\documentclass[aps,prl,showpacs,twocolumn,,superscriptaddress]{revtex4-1}

\usepackage{amsmath}
\usepackage{amsfonts}
\usepackage{amssymb}
\usepackage{epsfig}

\usepackage{graphicx}

\usepackage{hyphenat}
\usepackage{hyperref}
\usepackage{placeins}	
\usepackage{color}

\usepackage{array}
\newcolumntype{C}[1]{>{\centering\arraybackslash}m{#1}}
\usepackage{overpic}

\DeclareMathOperator{\Hamil}{\mathcal{H}}  
\newcommand{\eexp}{\mathrm{e}} 
\newcommand{\imag}{\mathrm{i}}

\begin{document}

\title{Critical behavior of quantum magnets with long-range interactions in the thermodynamic limit}
\author{Sebastian Fey}
\affiliation{Lehrstuhl f\"ur Theoretische Physik I, Staudtstra{\ss}e 7, Universit\"at Erlangen-N\"urnberg, D-91058 Erlangen, Germany}
\author{Kai Phillip Schmidt}
\affiliation{Lehrstuhl f\"ur Theoretische Physik I, Staudtstra{\ss}e 7, Universit\"at Erlangen-N\"urnberg, D-91058 Erlangen, Germany}

\begin{abstract}
Quasiparticle properties of quantum magnets with long-range interactions are investigated by high-order linked-cluster expansions in the thermodynamic limit. It is established that perturbative continuous unitary transformations on white graphs are a promising and flexible approach to treat long-range interactions in quantum many-body systems. We exemplify this scheme for the one-dimensional transverse-field Ising chain with long-range interactions. For this model the elementary Quasiparticle gap is determined allowing to access the quantum-critical regime including critical exponents and multiplicative logarithmic corrections for the ferro- and antiferromagnetic case.    
\end{abstract}

\maketitle

Correlated quantum many-body systems play an important role in various areas in modern physics, since fascinating quantum phases with exotic excitations as well as novel collective quantum behavior are expected. In many cases these correlations are induced by almost local interactions, e.g.,~the screened Coulomb interaction of the Hubbard model in correlated electron systems or the Ising and Heisenberg interactions between nearest neighbors in quantum magnetism. In contrast, there are many important physical systems with long-range interactions, which come more and more into focus \cite{Bitko1996,Chakraborty2004,Bramwell2001,Castelnovo2008,Mengotti2009,Lahaye2009,Peter2012,Britton2012,Islam2013,Jurcevic2014,Richerme2014,Mahmoudian2015,Bohnet1297}. One example for long-range interactions in condensed-matter physics are dipolar interactions between spins in so-called spin-ice materials giving rise to emergent magnetic monopoles \cite{Castelnovo2008}. Another important platform to engineer quantum many-body lattice models with long-range interactions are trapped cold ion systems in quantum optics for which the nature of interactions can be varied flexibly \cite{Britton2012, Islam2013}. Here an enormous experimental progress has been achieved over the last years allowing to realize one- and two-dimensional quantum-spin models and to investigate the properties of Quasiparticle excitations \cite{Britton2012,Jurcevic2014,Richerme2014,Bohnet1297}.    

Naturally, the theoretical treatment of long-range interactions in quantum many-body systems is notoriously complicated. This is especially true for the majority of numerical approaches which are usually applied to finite systems \cite{Laflorencie2005,Sandvik2010,Gorshkov2011,Koffel2012,Knap2013,Humeniuk2016} with a few exceptions like, for example, variational tensor network techniques. As a consequence, most investigations have focused on ground-state properties of one-dimensional quantum systems. One important tool to study quantum-lattice models directly in the thermodynamic limit and therefore avoiding finite-size effects are linked-cluster expansions (LCEs) which have been applied successfully in any dimension for models with short-range interactions in the past \cite{Gelfand96,Trebst2000,Knetter2000,Knetter2001,Oitmaa2006,Yang2010,Schulz2013,Coester2015}. Here the physical properties of the ground state and of Quasiparticle excitations  are determined via a full graph decomposition in topologically distinct graphs. However, the use of LCEs for systems with long-range interactions appears to be almost impossible, since the number of graphs in any order diverges due to the infinite number of different coupling constants. 

In this letter, we establish that this is not the case. LCEs up to high order in perturbation can be set up successfully by applying the recently developed white-graph expansion \cite{Coester2015}. Our approach is flexible, e.g.,~it can be used a priori in any spatial dimension as well as for arbitrary interactions including geometric frustration. As a proof of principle, we determine the Quasiparticle gap of the one-dimensional transverse-field Ising model (TFIM) with long-range interactions in the polarized high-field phase. This model has recently been realized in experiments on cold trapped ion systems \cite{Jurcevic2014,Bohnet1297} and is relevant for solid-state physics \cite{Bitko1996,Chakraborty2004}. Furthermore, our findings for the quantum-critical line can be compared to other numerical investigations \cite{Koffel2012,Knap2013} in order to gauge the quality of our approach. Finally, we extract the corresponding critical exponent from our LCEs. 

{\it{Set up:}}
We consider an Hamiltonian $\mathcal{H}$ at zero temperature of the form 
\begin{eqnarray}
\label{Eq:Hami}
{\mathcal H}&=&{\mathcal H}_0+\lambda\,\hat{V}\nonumber\\
            &=&{\mathcal H}_0+\lambda\sum_{i,j}{\mathcal V}\left[ g(i-j)\right] \quad ,
\end{eqnarray}
where ${\mathcal H}_0\equiv E_0+\sum_{i,\mu} \hat{f}^\dagger_{i,\mu}\hat{f}^{\phantom{\dagger}}_{i,\mu}$ is easily diagonalized in terms of supersites. In practice, a supersite might be a single spin, a dimer of two sites, or any collection of elementary sites which are suitable to describe the quantum phase under investigation. Here we assume that ${\mathcal H}_0$ has an equidistant spectrum with an energy gap $\Delta=1$ bounded from below by $E_0$. The lowest energy of a single supersite $E_0/N$ with $N$ being the number of supersites is considered to be nondegenerate (although degeneracies can be treated similarly with our approach). The sum over $\mu$ runs over all excited local degrees of freedom of a single supersite, e.g.,~for a single spin 1/2 there is only one local excitation corresponding to a local spin flip while for an antiferromagnetically coupled dimer of two spins 1/2 there are three degenerate local triplet excitations. The long-range interaction ${\mathcal V}\left[ g(i-j)\right]$ couples different supersites $i$ and $j$ so that $g(i-j)$ denotes the coupling strength. Here we concentrate on two-supersite interactions and a single parameter $\lambda$, but generalizations are straightforward.

The unperturbed ground state $|{\rm ref}\rangle$ at $\lambda=0$ with energy $E_0$ is interpreted as the vacuum and is given as the product state \mbox{$|{\rm ref}\rangle\equiv |0\rangle \cdots |0\rangle $} with $|0\rangle$ being the ground state of a supersite. Local excitations of type $\mu$ on supersite $i$ are created by $\hat{f}^\dagger_{i,\mu}|{\rm ref}\rangle$. It is always possible to introduce the counting operator \mbox{$\mathcal{Q}\equiv\sum_i\hat{n}_i\equiv \sum_{i,\mu} \hat{f}^\dagger_{i,\mu}\hat{f}^{\phantom{\dagger}}_{i,\mu}$} and to write ${\mathcal H}_0\equiv E_0+\mathcal{Q}$.
  
The Hamiltonian \eqref{Eq:Hami} can then be expressed as
\begin{equation}
\label{Eq:Hami_final}
{\cal H}={\mathcal H}_0+ \sum_{n=-N_{\rm max}}^{N_{\rm max}} \hat{T}_n \quad ,
\end{equation}
where $\lambda\hat{V}\equiv\sum_n\hat{T}_n$ and $[\mathcal{Q},\hat{T}_n]=n\hat{T}_n$. The operator \mbox{$\hat{T}_n \equiv\sum_{i,j} g(i-j) \hat{\tau}_n^{ij}$} corresponds to all operators where the change of energy quanta with respect to $\mathcal{Q}$ is exactly~$n$. Note that we have included $\lambda$ in the definition of the operators $\hat{\tau}_n^{ij}$ involving the supersites $i$ and $j$. The maximal (finite) change in energy quanta is called~$\pm N_{\rm max}$. 

{\it{Approach:}} Hamiltonians \eqref{Eq:Hami_final} can be well treated by the method of perturbative continuous unitary transformations (pCUTs) \cite{Knetter2000} and, more specifically, by the recently introduced white-graph expansion \cite{Coester2015}. Here this approach is extended to long-range interactions ${\mathcal V}\left[ g(i-j)\right]$.
 
In pCUTs, Hamiltonian \eqref{Eq:Hami_final} is mapped model-independently up to high orders in perturbation to an effective Hamiltonian $\mathcal{H}_\text{eff}$ with $[\mathcal{H}_{\rm eff},\mathcal{Q}]=0$. The general structure of $\mathcal{H}_{\rm eff}$ is then a weighted sum of operator products $\hat{T}_{\nu_1}\cdots \hat{T}_{\nu_k}$ in order $k$ perturbation theory, where $\hat{T}_{\nu_j}$ are from the pool of $\hat{T}_n$ in Eq.~\eqref{Eq:Hami_final} for each $j\in\{1,\ldots,k\}$. The block-diagonal $\mathcal{H}_\text{eff}$ conserves the number of Quasiparticles (qp). This represents a major simplification of the quantum many-body problem, since one can treat each Quasiparticle block, corresponding only to a few-body problem, separately. Physically, the zero Quasiparticle sector contains the ground-state energy of the system whereas the one Quasiparticle block gives access to the one Quasiparticle dispersion and therefore to the one-qp gap. Higher Quasiparticle blocks represent interacting few-body quantum systems.       

\begin{figure} [t!]
 \includegraphics[width=0.45\textwidth]{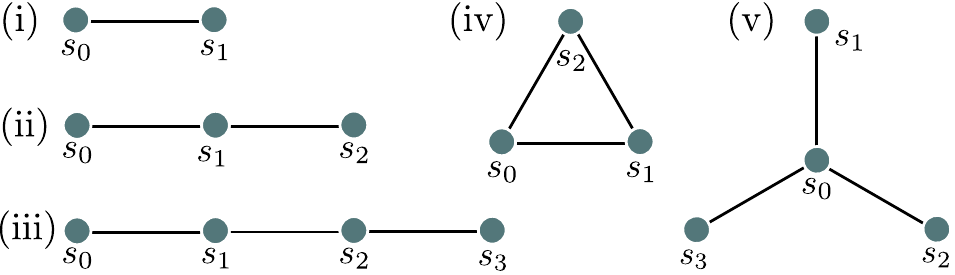}
 \caption{Illustration of all white graphs up to three links necessary for order three perturbation theory. Circles denote supersites $s_\nu$ while lines correspond to interactions ${\mathcal V}\left[ g(s_{\nu_1}-s_{\nu_2})\right]$ linking two supersites $s_1$ and $s_2$ on the graphs due to the interaction $g$. These white graphs have to be embedded into the system in the thermodynamic limit by identifying supersites $s_{\nu}$ of the graphs with the actual supersites $i$ of the lattice. For a long-range interaction $g(i-j)$ there are infinitely many embeddings for each graph.}
 \label{fig:graphs_u2}
\end{figure}

The more demanding part in pCUTs is model-dependent and corresponds to a normal-ordering of $\mathcal{H}_\text{eff}$. This is most efficiently done via a full graph decomposition in linked graphs using the linked-cluster theorem and an appropriate embedding scheme afterwards. In order $k$ perturbation theory, only linked graphs up to $k$ links have to be considered (see also Fig.~\ref{fig:graphs_u2}). A link between supersites $i$ and $j$ is introduced by the interaction ${\mathcal V}\left[ g(i-j)\right]=\sum_n g(i-j) \hat{\tau}_n^{ij}$, where each coupling $g(i-j)$ is associated with a different ``color''. In contrast to short-range interactions with only one (or a few) number of different colors, there are infinitely many different colors already in first-order perturbation theory for long-range interactions and the usual LCEs break down. 

At this point the recently introduced white-graph expansion \cite{Coester2015} turns out to be extremely useful. The essential idea is not to fix colors on graphs in advance, but to keep all relevant information during the calculation on graphs, so that one has to re-introduce colors only at the end of the calculation during the final embedding procedure. In the simplest realization, one introduces different parameters $\lambda_{j}$ on all $N_l$ links $l_j$ with $j\in\{1,\ldots,N_l\}$ of a given graph. The calculation then yields contributions proportional to $\lambda_{1}^{k_1}\cdots\lambda_{N_l}^{k_{N_l}}$ with $k_1+\cdots + k_{N_l}=k$ in order $k$ which have to be embedded in the infinite lattice by replacing the $\lambda_{j}$ by the function $g$. Note that also more sophisticated schemes are possible, which is a consequence of the fact that $\mathcal{H}_\text{eff}$ is given in second quantization and in the thermodynamic limit \cite{Coester2015}. 

Therefore, due to white graphs, it is not anymore the generation of and the calculation on graphs which is most challenging for LCEs with long-range interaction, but it is the final embedding procedure. Indeed, one obtains up to $k$ infinite sums in order $k$ perturbation theory for the different matrix elements of $\mathcal{H}_\text{eff}$. Physically, an infinite sum originates from the fact that each link of a given graph has to be embedded infinitely many times on the lattice due to the long-range nature of the interaction. The number of infinite sums then equals the number of different links of a graph, i.e.~one obtains maximally $k$ infinite sums for the case of the chain graph with $k$ different links. These infinite sums have to be evaluated quantitatively in order to capture the physical processes of the effective Hamiltonian properly. The technical details of this evaluation procedure are given in the Appendices \ref{App:A} to \ref{App:C}. We stress that the infinite sums are in general nested, since extra conditions have to be imposed when embedding graph sites on the lattice. Important examples are the chain graphs (i)-(iii) in Fig.~\ref{fig:graphs_u2}, where it is not allowed to embed two graph sites on the same lattice site.

Let us illustrate the appearance of infinite sums during the embedding process for the simplest graph (i) with one link as shown in Fig.~\ref{fig:graphs_u2}.  The interaction between two supersites $s_0$ and $s_1$ on this graph yields in first-order perturbation theory operators of the form \mbox{$g(s_0-s_1)\hat{\tau}_0^{s_0s_1}$} which can for example represent a nearest-neighbor hopping amplitude of a Quasiparticle. In the next step this white-graph contribution has to be embedded into the infinite lattice. Since the interaction is long-range, there are infinitely many, usually different embeddings of this graph. The final contribution of graph (i) in the thermodynamic limit then yields
\begin{align}
  \frac{1}{2}\sum_{\substack{\delta=-\infty \\ \delta \neq0}}^\infty g(\delta\equiv s_0-s_1) \,\hat{\tau}_0^{s_0s_1}\quad .
\end{align}
For a general graph, consisting of $n$ links, each link $l_j$ typically yields such an infinite sum over distances $\delta_{l_j}$. Additionally, if graphs contain loops, each loop introduces the extra condition $\sum_{l_j\in \{\rm loop\}}\delta_{l_j}=0$ on the involved distances of the loop.  

Such products of sums have to be evaluated efficiently in order to reach quantitative results up to high orders in perturbation. But apart from that, this expansion allows to formulate high-order LCEs for long-range interactions in quantum lattice models on general grounds.  

{\it{Application:}} As an important example we consider the transverse-field Ising chain with long-range interactions given by 
\begin{align}
  \Hamil =-\frac{1}{2}\sum_{j}\sigma_j^z -\lambda\sum_{i\neq j} \frac{1}{|i-j|^\alpha}\sigma_i^x\sigma_{j}^x\quad , \label{eq:H_tfim_orig}
\end{align}
where the sums run over the sites of the infinite chain, $\sigma^\kappa$ with $\kappa\in\{x,y,z\}$ denotes the Pauli matrices, and $\alpha$ varies from the short-range limit $\alpha\rightarrow\infty$ up to the ultra long-range case $\alpha=0$. Positive (negative) \mbox{$\lambda$} corresponds to ferromagnetic (antiferromagnetic) Ising interactions. 

Introducing hardcore boson operators $b^\dagger_j$, $b^{\phantom{\dagger}}_j$, and \mbox{$\hat{n}_j\equiv b^\dagger_j b^{\phantom{\dagger}}_j$} on site $j$ by applying the Matsubara-Matsuda transformation \cite{Matsubara1956} (see also Eq.~\eqref{eq::mama}), we can rewrite Eq.~\eqref{eq:H_tfim_orig} up to the constant $-N/2$ as
\begin{align}
  \Hamil =\sum_{j} \hat{n}_j -\lambda\sum_{i\neq j} g_\alpha(i-j)\left( b^\dagger_i b^\dagger_j + b^\dagger_i b^{\phantom{\dagger}}_j + {\rm H.c.}\right) \, , \label{eq:H_tfim_orig_boson}
\end{align}
which is indeed of the form \eqref{Eq:Hami_final} with $N_{\rm max}=2$ and \mbox{$g_\alpha(i-j)\equiv |i-j|^{-\alpha}$}.

This model possesses two gapped phases: a polarized phase for small $|\lambda|$ and a $\mathcal{Z}_2$ symmetry-broken ground state for large $|\lambda|$. We have applied the above formulated LCE to calculate the one-qp gap $\Delta_{\rm f/af}$ of the polarized phase in the high-field limit $|\lambda|\rightarrow 0$ up to order 8 in $\lambda$ for the ferromagnetic (f) / antiferromagnetic (af) case. 

The two leading orders can be evaluated exactly, since only the two chain graphs (i) and (ii) in Fig.~\ref{fig:graphs_u2} without loops are relevant. One obtains
\begin{align}
	\Delta_{\rm f} &= 1-2 \zeta (\alpha) \lambda  + 2 \left(\zeta (2 \alpha)-\zeta (\alpha)^2 \right) \lambda^2 +\mathcal{O}(\lambda^3)\\
	\begin{split}
		\Delta_{\rm af} &=1 +\left(2^{1-\alpha} \left(2^{\alpha}-2\right) \zeta (\alpha)\right) \lambda \,+\\
			& \left(2 \zeta (2 \alpha)-2^{1-2 \alpha} \left(2^{\alpha}-2\right)^2 \zeta (\alpha)^2\right) \lambda^2 +\mathcal{O}(\lambda^3 )
	\end{split}
\end{align}
where $\zeta(\alpha)$ is the Riemann zeta function. The higher orders of the gap are determined by summing the various infinite sums using finite limits $\mathcal{N}$ and performing appropriate extrapolations of the numerical data sequences as outlined in Appendix \ref{App:B}. Apart from the Wynn algorithm \cite{Wynn1956}, we used a scaling in $1/\mathcal{N}^{\alpha-1}$ ($1/\mathcal{N}^{\alpha}$) for f (af) Ising interactions. This scaling can be derived analytically for any product of Riemann zeta functions and is the correct scaling for every coefficient of the gap series (see Appendices \ref{App:B} and \ref{App:C}). Both extrapolation schemes give consistent results, but the scaling works generically better so that we display these results below.

\begin{figure} [t!]
 \includegraphics[width=0.85\columnwidth]{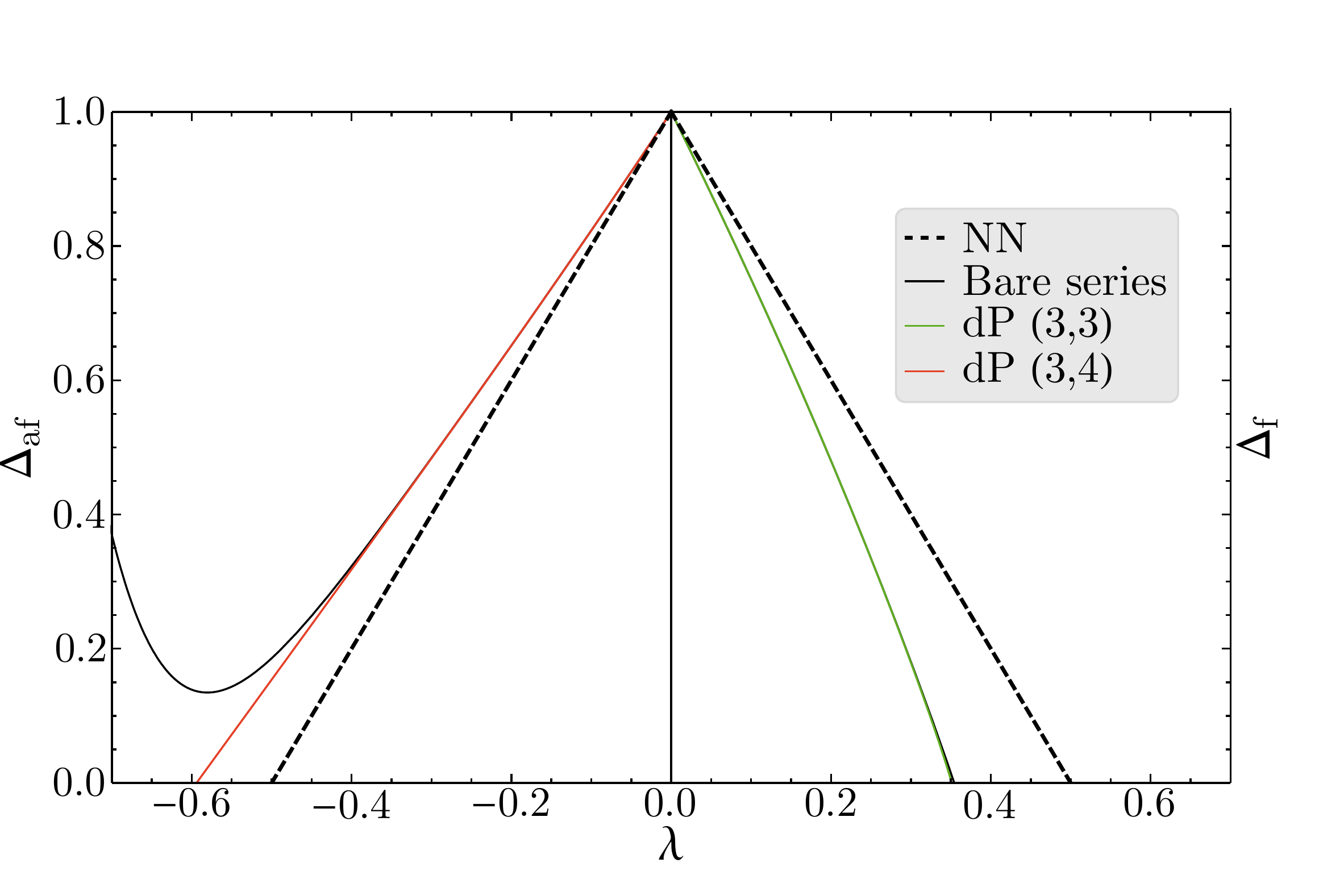}
  \caption{The one-qp gap $\Delta_{\rm f/af}$ as a function of $\lambda$ for f (af) Ising interactions with exponent $\alpha=3$. Solid black lines correspond to the bare order-8 series, while other solid lines refer to representative DlogPad\'e extrapolants. Dashed black lines are the exact one-qp gaps for the nearest-neighbor TFIM in the limit $\alpha\rightarrow\infty$.}
 \label{fig:gap_alpha_3}
\end{figure}

{\it{Ferromagnetic case:}}
Let us focus on ferromagnetic interactions $\lambda>0$. Here only exponents $\alpha>1$ are well defined. In our LCE this becomes apparent due to divergencies in the infinite sums for $\alpha\leq 1$. In the opposite limit $\alpha\rightarrow\infty$ one recovers the exact solution of the nearest-neighbor TFIM \mbox{$\Delta_{\rm f}=1-2\lambda$} yielding a quantum phase transition between the polarized phase and the symmetry-broken phase at $\lambda_{\rm c}=0.5$ with an exponent $z\nu=1$. Any ferromagnetic long-range interaction with finite $\alpha$ stabilizes the symmetry-broken phase and one expects $\lambda_{\rm c}<0.5$. This is illustrated in Fig.~\ref{fig:gap_alpha_3} for $\alpha=3$.

\begin{figure} [t!]
 \includegraphics[width=0.85\columnwidth]{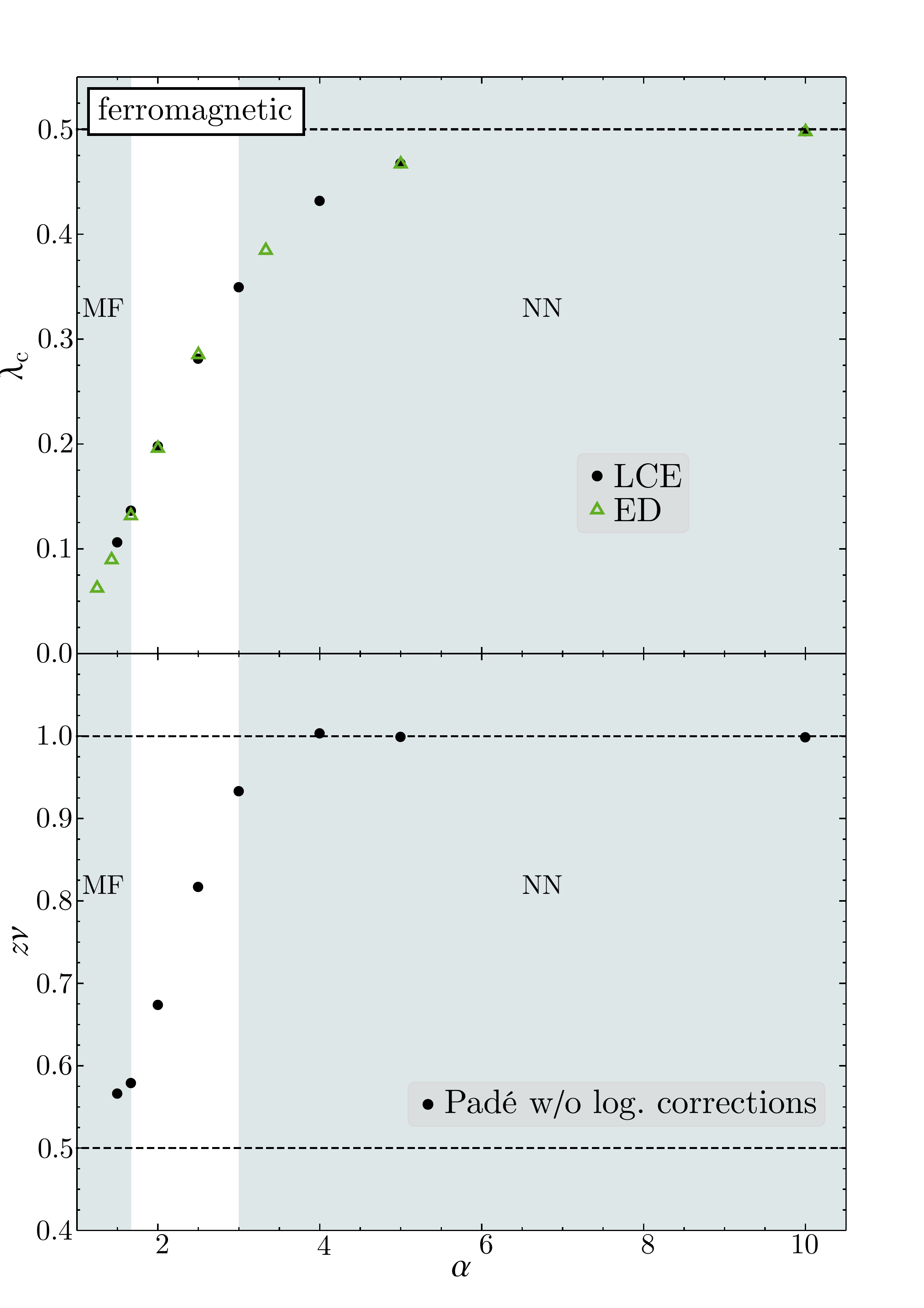}
 \caption{Quantum-critical points $\lambda_{\rm c}$ (upper panel) and critical exponents $z\nu$ (lower panel) as a function of $\alpha$ for the ferromagnetic case. Black circles represent averaged DlogPad\'e extrapolants of $\Delta_{\rm f}$ of the highest available order. Green triangles correspond to scaled ED data from Ref.~\onlinecite{Knap2013}. The mean-field (MF) and nearest-neighbor (NN) TFIM universality classes are illustrated as grey backgrounds and the associated critical exponents as horizontal dashed lines.}
 \label{fig:ferro_val}
\end{figure}

We use DlogPad\'e extrapolation of the gap series \cite{Guttmann1989} to estimate the quantum critical points $\lambda_{\rm c}$ for various values of $\alpha$ (see also Appendix \ref{App:D}). The results are displayed together with scaled exact diagonalization (ED) data from Ref.~\onlinecite{Knap2013} in Fig.~\ref{fig:ferro_val}. One obtains very good agreement between both approaches for a wide range of $\alpha$ values. Only for the demanding regime of small $\alpha$ visible deviations can be seen. Here the extrapolation of the series as well as the finite-size scaling of ED data becomes challenging.  

Next we turn to the nature of the quantum phase transition as a function of $\alpha$. From one-loop renormalization group calculations \cite{Dutta2001,Knap2013}, one expects three different domains: i) the system is in the same universality class as the nearest-neighbor TFIM with $z\nu=1$ for $\alpha\geq 3$, ii) the system displays mean-field behavior $z\nu=1/2$ for $\alpha\leq 5/3$, and iii) the system has nontrivial continuously varying critical exponents for $5/3<\alpha<3$.  

We extracted the critical exponent $z\nu$ as a function of $\alpha$ from the DlogPad\'e extrapolation of $\Delta_{\rm f}$ which is shown in Fig.~\ref{fig:ferro_val}. As expected, the critical exponent is close to $1$ for $\alpha\geq 3$ and then continuously decreases for smaller values of $\alpha$. One should stress that any LCE is not able to resolve abrupt changes of critical exponents, since only finite orders enter into the extrapolation of the series. 

However, the visible deviation around \mbox{$\alpha=5/3$} is unexpected but can be traced back to the presence of multiplicative logarithmic corrections at the ``upper critical $\alpha$'' similar to the upper critical dimension $d=3$ for the nearest-neighbor TFIM. For the latter one finds $p=-1/6$ for $d=3$ from perturbative RG and series expansions \cite{Larkin1969,Brezin1973,Wegner1973,Zheng1994,Coester2016}. In our case, fixing $\lambda_{\rm c}=0.1374$ and $z\nu=1/2$, we find $p\approx -0.20(4)$ for \mbox{$\alpha=5/3$} when averaging over order-8 DlogPad\'e extrapolations. We stress that multiplicative logarithmic corrections are very sensitive on $\lambda_{\rm c}$. The extracted value for $p$ is therefore remarkably close to $-1/6$. This fully supports the idea that the quantum critical behavior induced by the long-range Ising interaction can effectively be understood in terms of the nearest-neighbor TFIM in an effective spatial dimension $d_{\rm eff}$.  

{\it{Antiferromagnetic case:}} The antiferromagnetic long-range TFIM behaves fundamentally different to the ferromagnetic case, which is mainly due to geometric frustration. As a consequence, any finite value of $\alpha$ enlarges (reduces) the polarized (symmetry-broken) phase compared to the nearest-neighbor TFIM for $\alpha\rightarrow\infty$. This is illustrated for $\alpha=3$ in Fig.~\ref{fig:gap_alpha_3}. In Ref.~\onlinecite{Koffel2012}, this phase diagram has been calculated by variational matrix product states (MPS). They found that the critical point increases monotonously from $\lambda_{\rm c}=-0.5$ to $\lambda_{\rm c}\rightarrow -\infty$ when varying $\alpha$ from $\infty$ to $0$.  

\begin{figure} [t!]
 \includegraphics[width=0.85\columnwidth]{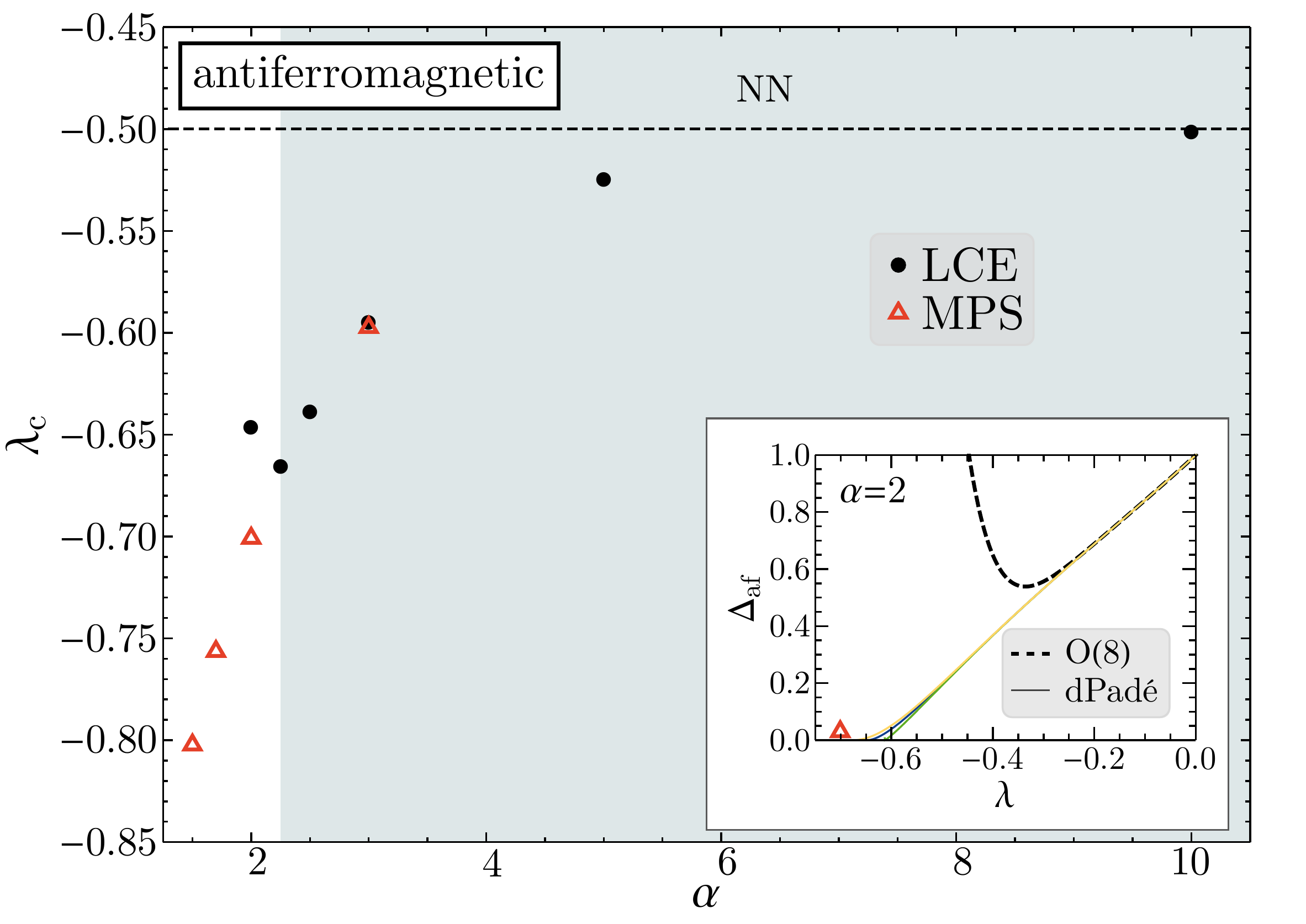}
 \caption{Quantum-critical points $\lambda_{\rm c}$ as a function of $\alpha$ for the antiferromagnetic case. Black circles represent averaged DlogPad\'e extrapolants of $\Delta_{\rm af}$. Red triangles correspond to MPS data from Ref.~\onlinecite{Koffel2012}. The nearest-neighbor (NN) TFIM universality class is illustrated with a gray background. {\it Inset}: The gap $\Delta_{\rm af}$ as a function of $\lambda$ for $\alpha=2$. Dashed line refers to bare series and solid lines correspond to different order-8 DlogPad\'e extrapolants.}
 \label{fig:antiferro_val}
\end{figure}

We used DlogPad\'e extrapolation of $\Delta_{\rm af}$ to extract the critical point $\lambda_{\rm c}$ (see Fig.~\ref{fig:antiferro_val}) and the critical exponent $z\nu$ for various values of $\alpha$. From renormalization group calculations one expects the system to be in the same universality class as the nearest-neighbor TFIM for~$\alpha\geq 9/4$ \cite{Koffel2012}. Our LCE for the critical line are in quantitative agreement with MPS calculations in this $\alpha$-regime and we find indeed a critical exponent $z\nu$ close to one, e.g.,~$z\nu=1.012(3)$ for $\alpha=9/4$. The situation is more peculiar for $\alpha<9/4$. Here the MPS calculations suggests continuously varying critical exponents and, furthermore, a breakdown of the area law due to the long-range nature of the interaction even inside the gapped polarized phase \cite{Koffel2012,Peter2012}. Interestingly, the deviations LCE and MPS are already large for $\alpha=2$ (see inset of Fig.~\ref{fig:antiferro_val}). This suggests that either the critical exponent $z\nu$ grows extremely for $\alpha<9/4$ (we find $z\nu=1.7(5)$ for $\alpha=2$), the quantum-critical breakdown of the polarized phase is not at all described by a simple algebraic divergence, but nonperturbative terms are present which cannot be captured by the LCE, or this highly entangled and long-range $\alpha$-regime is also very challenging for the MPS calculation.     
  
{\it{Conclusion:}} We established that LCEs using perturbative continuous unitary transformations are a flexible and promising approach to treat long-range interactions in quantum many-body systems. As a proof of principle, we have applied LCEs to the long-range transverse-field Ising chain obtaining highly competitive results compared to existing numerical data. This opens the door for microscopic calculations of two- and three-dimensional correlated quantum systems with long-range interactions of arbitrary nature important for condensed matter physics and quantum optics. 

{\it{Acknowledgement:}}
We thank Kris Coester and Michael Knap for fruitful discussions as well as Michael Knap and Luca Tagliacozzo for providing us with their numerical data.

%
%
\appendix

\section{White-graph expansion of the long-range TFIM}
\label{App:A}
We investigated the critical behavior of the one-dimensional TFIM with algebraically decaying long-range interactions
\begin{align}
  \Hamil =-\frac{1}{2}\sum_{j}\sigma_j^z -\lambda\sum_{i\neq j} \frac{1}{|i-j|^\alpha}\sigma_i^x\sigma_{j}^x\quad . \label{eq_supp:H_tfim_orig}
\end{align}
using perturbative continuous unitary transformations about the high-field limit.

To this end we perform a Matsubara-Matsuda transformation \cite{Matsubara1956} and replace the Pauli matrices $\sigma_i^\kappa$, $\kappa\in\{x,z\}$ with hardcore-boson annihilation (creation) operators $b_i^{(\dagger)}$
\begin{align}
 \label{eq::mama}
	\sigma_i^x = b_i^\dagger+b_i, && \sigma_i^z=1-2\hat{n}_i,\quad \text{with}~\hat{n}_i=b_i^\dagger b_i^{\phantom{\dagger}}\quad .
\end{align}

The ground state of polarized spins in the limit $\lambda\rightarrow 0$ becomes the vacuum state in the bosonic Quasiparticle picture while spin-flip excitations correspond to hardcore bosons located on the lattice sites. In this formulation we end up with Eq.~(5) in the main body of the manuscript
\begin{align}
  \Hamil =\sum_{j} \hat{n}_j -\lambda\sum_{i\neq j} g_\alpha(i-j)\left( b^\dagger_i b^\dagger_j + b^\dagger_i b^{\phantom{\dagger}}_j + {\rm H.c.}\right) \, , \label{eq_supp:H_tfim_orig_boson}
\end{align}
which is of the form (2) with $N_{\rm max}=2$ and \mbox{$g_\alpha(i-j)\equiv |i-j|^{-\alpha}$}.

In pCUTs, Hamiltonian (2) is mapped up to high orders in perturbation to an effective Hamiltonian $\mathcal{H}_\text{eff}$ with $[\mathcal{H}_{\rm eff},\mathcal{Q}]=0$. The block-diagonal $\mathcal{H}_\text{eff}$ conserves therefore the number of Quasiparticles which correspond to dressed spin-flip excitations in our case. Here we focus on the one-qp sector where the effective Hamiltonian is given as a hopping Hamiltonian of the form
\begin{align}
 \Hamil^{\rm 1qp} =\sum_{i}\sum_\delta a_\delta \left( b^\dagger_i b^{\phantom\dagger}_{i+\delta} + {\rm H.c.}\right) \, , \label{eq_supp:H_eff_1qp}
\end{align}
with $a_\delta$ denoting the hopping amplitude of distance $\delta$ between two sites on the chain. In pCUTs, these hopping amplitudes are derived up to high orders in perturbation. 

Using the Fourier transformation
\begin{align}
	b_{j}^\dagger = \frac{1}{\sqrt{N_{\mathrm{s}}}}\sum\limits_q \eexp^{\imag q j}b_{q}^\dagger,\quad
	b_{j}^{\phantom\dagger} &= \frac{1}{\sqrt{N_{\mathrm{s}}}}\sum\limits_q \eexp^{-\imag q j}b_{q}^{\phantom\dagger }
\end{align}
with the number of lattice sites $N_{\mathrm{s}}$, the one-qp Hamiltonian \eqref{eq_supp:H_eff_1qp} is readily diagonalized  
\begin{align}
\Hamil^{\rm 1qp}  = \sum\limits_{q} \omega_q \,b_{q}^\dagger b_{q}^{\phantom\dagger}\quad .
\end{align}
Here $\omega_q=a_0+2\sum_{\delta>0} a_\delta \cos(q\,\delta)$ is the one-qp dispersion. The minimum of the dispersion corresponds to the one-qp gap $\Delta\equiv {\rm min}_q\,\omega_q$. For the long-range TFIM the one-qp gap $\Delta$ is located at momentum $q_{\Delta}=0$ for a ferromagnetic and $q_\Delta=\pi$ for an antiferromagnetic Ising interaction, respectively.

We have calculated this Quasiparticle gap $\Delta$ as a series in the perturbation parameter $\lambda$
\begin{align}
\Delta (\lambda) = 1 + p_1 \lambda + p_2 \lambda^2 +\dots~+ p_{k} \lambda^{k} \label{eq_supp:gap_series_general}
\end{align}
up to order $k=8$. All prefactors $p_r$ depend on $q_{\Delta}$ and can be analytically expressed as
\begin{align}
	p_r=\sum_{\gamma} t_{r,\gamma}
\end{align}	
where the sum runs over all graphs $\gamma$ contributing to the given order $r$ (c.f. Fig.~(1) in the main body of the text for an overview of all graphs up to order 3). In order 8 there are 358 graphs in total.

The parameter $t_{r,\gamma}$ is the unique contribution of graph $\gamma$ to the coefficient $p_r$ in which the aforementioned infinite sums appear due to the embedding process. In practice, we introduce a different coupling $\lambda_{j}$ for each link $l_j$ of a given graph $\gamma$. The pCUT calculation in order $r$ then yields hopping amplitudes between sites $\nu$ and $\nu+\delta$ of the form 
\begin{align}
 \sum_{\{r_j\}} A_{\nu,\nu+\delta,\gamma}\left({\{r_j\}}\right)\,\lambda_{1}^{r_1}\cdots\lambda^{r_{\rm max}}_{{\rm max}}
\end{align}
where $\sum_{j}r_j=r$ holds for each summand and the coefficients $A_{\nu,\nu+\delta,\gamma}\left({\{r_j\}}\right)$ are exact fractions. In the next step one has to embed the graph links $l_j$ into the infinite chain which implies
\begin{align}
 \lambda_{j}^{r_j} \longrightarrow -\lambda^{r_j} \left(\frac{1}{|\delta_{l_j}|^{\alpha}}\right)^{r_j}
\end{align}
and summing over all possible embeddings of graph $\gamma$. 

Fourier transformation of all hopping processes yields the parameter $t_{r,\gamma}$ which can be written for general momentum $q$ as
\begin{align}
 \label{eq:t_r_gamma}
	t_{r,\gamma} = a_{0,\gamma}^{(r)} + 2 \sum_{\substack{\delta\in\gamma\\\delta>0}} a_{\delta,\gamma}^{(r)} \cos(q\,\delta)\quad ,
\end{align}
where
\begin{align}
	a_{\delta,\gamma}^{(r)} =  \xi_\gamma &\sum_{\nu<N_\gamma}\sum_{\{r_j\}}  A_{\nu,\nu+\delta,\gamma}\left({\{r_j\}}\right)\,\notag\\ & ~\sum_{s_{N_\gamma}}\dots\sum_{s_2}\sum_{s_1}\, f_{\nu,\nu+\delta,\gamma}^{{\{r_j\}}}(\{s_j\})\quad .
\end{align}
Here $\nu=0..(N_\gamma-1)$ where $N_\gamma$ is the total number of the graph's lattice sites, $\xi_\gamma$ is a factor compensating the overcounting in the summation due to the graph symmetry, and $A_{\nu,\nu+\delta,\gamma}\left({\{r_j\}}\right)$ is the pCUT graph-dependent hopping amplitude from graph-site $\nu$ to $\nu+\delta$. The lattice-site indices on the infinite chain are denoted by $s_\nu$. For the local hopping $a_{0,\gamma}^{(r)}$ the graph's ground-state energy is subtracted from the one-qp energy. The factor $f_{\nu,\nu+\delta,\gamma}^{{\{r_j\}}}(\{s_j\})$ is a graph-dependent product of fractions arising from the long-range interactions
\begin{align}
	f_{\nu,\nu+\delta,\gamma}^{{\{r_j\}}}(\{s_j\})=\lambda^r \prod_{\{r_j\}}\frac{1}{\left|s_{\nu_j}-s_{\nu'_j}\right|^{r_j\alpha}}
\end{align}	
where the sum over all $r_m$ equals the order $r$ and \mbox{$s_{\nu_j}-s_{\nu'_j}=\delta_{l_j}\neq 0$}.

\begin{figure} [t!]
 \includegraphics[width=0.45\textwidth]{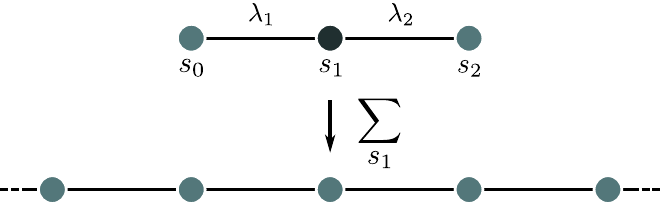}
 \caption{Embedding of a graph with three sites $s_\nu$ into the one-dimensional lattice in the thermodynamic limit. One after the other, each of these sites have to be set to any of the (still unoccupied) lattice sites to get the contribution of all the realizations of the graph in the actual lattice.}
 \label{fig:embedding}
\end{figure}

As a simple example, which arises from the pCUT calculation in order $r=3$, let us consider graph~(ii) in Fig.~(1) in the main body of the manuscript denoted from now on by $\gamma_{\rm (ii)}$. This chain graph has three sites $s_0$, $s_1$, and $s_2$ and two links $l_1$ between the first two sites and $l_2$ between the last two sites. Here we focus on a specific nearest-neighbor hopping between site $s_0$ to $s_1$ with a certain set of $\{r_j\}$ in order to illustrate the embedding procedure and we want to calculate in the following contribution of this process to the parameter $t_{3,\gamma_{\rm (ii)}}$.

The corresponding contribution for that hopping on graph $\gamma_{\rm (ii)}$ is given as
\begin{align}
	-\frac{1}{4}\lambda_{1}\lambda_{2}^2\quad .
\end{align}
The embedding process, illustrated in Fig.~\ref{fig:embedding}, means a summation over all possible realizations of that graph on the actual lattice. For a long-range interaction there are clearly infinitely many possibilities. In our example we get after embedding the following contribution to the parameter $t_{3,\gamma_{\rm (ii)}}$
\begin{align}
 \label{eq:embedding}
	\frac{1}{4}\lambda^3 \sum_{ \substack{ \delta_{l_2} =-\infty \\ \delta_{l_2}\neq -\delta_{l_1} \\ \delta_{l_2} \neq 0  }}^\infty \sum_{ \substack{ \delta_{l_1}=-\infty \\ \delta_{l_1} \neq 0  }}^\infty \frac{1}{|\delta_{l_1}|^\alpha}\frac{1}{|\delta_{l_2}|^{2\alpha}}\cos(q\delta_{l_1})
\end{align}
where the factor $\xi_{\gamma_{\rm (ii)}}=1/2$ comes from the graph's symmetry and accounts for a double counting of each realization of the graph on the lattice. This factor is canceled with the factor $2$ in Eq.~\eqref{eq:t_r_gamma}. The conditions $\delta_{l_2}\neq -\delta_{l_1}$ and $\delta_{l_j}\neq 0$ in the sums ensure that the possibility of two graph sites being located on the same lattice site is excluded.

For a quantitative evaluation of this expression the infinite sums still need to be calculated. This task proves to be difficult for a general value of $q$. Here we are only interested in the two specific momenta $q=0$ and $q=\pi$. In both cases expression \eqref{eq:embedding} can be evaluated analytically to a product of two Riemann zeta functions. For the ferromagnetic case $q=0$ one obtains 
\begin{align}
  \left(2\lambda^3\zeta(\alpha)\zeta(2\alpha)-1\lambda^3\zeta(3\alpha)\right)
\end{align}
and for the antiferromagnetic case one finds
\begin{align}
 \lambda^3 \left( 2 \left(2^{1-2 \alpha}-1\right) \zeta (\alpha) \zeta (2 \alpha) +2^{-3 \alpha} \left(8^\alpha-2\right) \zeta (3 \alpha) \right) \, .
\end{align}

\section{Extrapolation of data sequences}
\label{App:B}
The nested infinite sums appearing at perturbative orders $r>2$ cannot be evaluated analytically. Therefore we have calculated the various contributions by cutting the sums at finite limits $\mathcal{N}$. In this situation one has to find proper schemes to extrapolate the data sequences for different $\mathcal{N}$ to $\mathcal{N}\rightarrow\infty$. In practice we have applied the Wynn algorithm and performed proper scalings in $1/\mathcal{N}$ to the coefficients $p_r$ of the one-qp gap. We haven chosen to extrapolate the $p_r$ to minimize the number of extrapolations which have to be done in order to obtain~$\Delta$. 

We found that the behavior of the ferromagnetic data sequences is fundamentally different from the antiferromagnetic ones. The ferromagnetic sequences converge monotonically for large enough $\mathcal{N}$ while in the antiferromagnetic case one observes an alternating behavior about the exact value at $\mathcal{N}\rightarrow\infty$. As a consequence, the antiferromagnetic coefficients $p_r$ converge faster with $\mathcal{N}$ than the ferromagnetic parameters and the scaling behavior of both cases is different.   

\subsection{Wynn algorithm}
The sums are evaluated for fixed values of $\alpha$ as partial sums up to the upper boundary $\mathcal{N}$. In the antiferromagnetic case the partial sums are alternating. Therefore we consider only every second data point to get a monotonically converging series of data points (see also next section). These data points are extrapolated using Wynn's epsilon method \cite{Wynn1956}. Several extrapolations using a subset of the full series of points from $S_1$ up to $S_{\mathcal{N}}$ are made for each $p_r$. These are shown as red crosses in the figures. Afterwards the Wynn results are averaged using the best converged data points which is marked by a vertical black line in the figures (see e.g.,~Fig. \ref{fig:ext_kpi_a1.5}).

Wynn's epsilon method is an acceleration method for series which are converging slowly, as is the case especially for small values of $\alpha$. Setting the start values of the algorithm to $\epsilon_0(S_n)=S_n$ and $\epsilon_{-1}(S_n)=0$ the iteration reads
\begin{align}
	\epsilon_{k+1}(S_n)=\epsilon_{k-1}(S_{n+1})+\frac{1}{\epsilon_k(S_{n+1})-\epsilon_k(S_n)}\quad.
\end{align}

\subsection{Scaling}

As discussed above, each coefficient $p_r$ of the gap is a sum of various nested infinite sums. Truncating the infinite sums at a finite limit $\mathcal{N}$, one might wonder how the coefficients $p_r$ scale to the infinite-sum limit for different $\alpha$. Here we argue that each term of infinite sums scales similarly to the scaling of a product of Riemann zeta functions, which can be derived analytically and is therefore used as the proper scaling of the numerical data sequences.

\subsubsection{ferromagnetic case}
 
If one sets $q=0$ in the coefficients $p_r$ relevant for ferromagnetic Ising interactions, then all infinite sums become monotonic (see for example Eq.~\eqref{eq:embedding}). We therefore start by considering a single harmonic sum of the form
\begin{align}
 \sum_{\delta=1}^{\mathcal{N}} \frac{1}{\delta^\alpha}
\end{align}
which converges to the Riemann zeta function $\zeta (\alpha)$ for $\mathcal{N}\rightarrow\infty$. We are interested in the leading asymptotics for large $\mathcal{N}$ of the full sum, i.e.~we consider the difference
\begin{align}
 \sum_{\delta=\mathcal{N}+1}^{\infty} \frac{1}{\delta^\alpha}= \zeta(\alpha)-\sum_{\delta=1}^{\mathcal{N}} \frac{1}{\delta^\alpha}\quad .
\end{align}

We therefore replace the sum by an integral and find for large $\mathcal{N}$ and $\alpha>1$
\begin{align}
 \int_{\mathcal{N}+1}^\infty {\rm d}\,\delta\,\frac{1}{\delta^\alpha} = \frac{(\mathcal{N}+1)^{-\alpha+1}}{-\alpha+1}\propto \frac{\mathcal{N}^{-\alpha+1}}{-\alpha+1}\quad .
\end{align}
In the coefficients $p_r$ there are sums of terms with a different number of infinite sums. If these sums are independent, then one can factorize them and obtains generically a product of harmonic sums of the form 
\begin{align}
 \left(\sum_{\delta_1=1}^{\mathcal{N}} \frac{1}{\delta_1^\alpha}\right)\left(\sum_{\delta_2=1}^{\mathcal{N}} \frac{1}{\delta_2^\alpha}\right)\cdots \left(\sum_{\delta_m=1}^{\mathcal{N}} \frac{1}{\delta_m^\alpha}\right)\, .
\end{align}
Each terms scales for large $\mathcal{N}$ as $\zeta(\alpha)+\frac{\mathcal{N}^{-\alpha+1}}{-\alpha+1}$ so that the leading scaling of the product is
\begin{align}
 \zeta(\alpha)^m+m\zeta(\alpha) \frac{\mathcal{N}^{-\alpha+1}}{-\alpha+1} + \ldots \quad .
\end{align}
So all products scale with the same exponent $(1-\alpha)$ independent of $m$ which we also confirmed numerically. In the following we used this scaling for the coefficients $p_r$ of the gap. Here we assume that the nested conditions in the sum, which usually spoil the possibility to factorize the sums, do not alter the scaling behavior. First, one can rewrite a nested product of sums often as a sum of unnested sums. Second, the term with the largest number of sums arises always from the longest chain graph contributing in a given order and the contribution of this chain graph contains always the factorized product of independent sums.

\subsubsection{antiferromagnetic case}
 
If one sets $q=\pi$ in the coefficients $p_r$ relevant for antiferromagnetic Ising interactions, then all infinite sums become alternating (see for example Eq.~\eqref{eq:embedding}). We therefore start by considering a single sum of the form
\begin{align}
 \sum_{\delta=1}^{\mathcal{N}}\, (-1)^\delta\, \frac{1}{\delta^\alpha}
\end{align}
and we denote the limiting value of the sum as $\epsilon (\alpha)$ for $\mathcal{N}\rightarrow\infty$. We are again interested in the leading asymptotics for large $\mathcal{N}$ of the full sum, i.e.~we consider the difference
\begin{align}
 \sum_{\delta=\mathcal{N}+1}^{\infty} (-1)^\delta\, \frac{1}{\delta^\alpha} = \epsilon(\alpha)-\sum_{\delta=1}^{\mathcal{N}} (-1)^\delta\, \frac{1}{\delta^\alpha}\quad .
\end{align}
We then separate odd and even orders corresponding to negative and positive contributions and we assume $N$ to be even
\begin{align}
 \sum_{\delta=\mathcal{N}+1}^{\infty}\, (-1)^\delta\, \frac{1}{\delta^\alpha}=\sum_{\delta=\frac{\mathcal{N}}{2}+1}^{\infty} \left( \frac{1}{(2\delta)^\alpha} - \frac{1}{(2\delta-1)^\alpha}\right) \quad .
\end{align}
This sum is again monotonic as above for the ferromagnetic case. The involved $\delta$ are large, since $\mathcal{N}$ is supposed to be large. We therefore perform the Taylor expansion $1/(2\delta-1)^\alpha\approx 1/(2\delta)^\alpha (1+\alpha/2\delta+\ldots )$ for the second term so that the sum is taken over $\alpha/(2\delta)^{\alpha+1}$. In the next step we replace the sum again by an integral and find the following scaling behavior
\begin{align}
 \int_{\frac{\mathcal{N}}{2}+1}^\infty {\rm d}\,\delta \frac{\alpha}{(2\delta)^{\alpha+1}} = -\alpha\frac{\left(\frac{\mathcal{N}}{2}+1\right)^{-\alpha}}{2^{\alpha+1}} \propto -\frac{\alpha}{2}\mathcal{N}^{-\alpha} \quad .
\end{align}
As for the ferromagnetic case, this can be generalized for products of independent sums to\begin{align}
 \epsilon(\alpha)^m-m\frac{\alpha}{2}\epsilon(\alpha)\,\,\mathcal{N}^{-\alpha}+\ldots \quad ,
\end{align}
where $\epsilon(\alpha)$ denotes the exact value for $\mathcal{N}\rightarrow\infty$. So all products scale with the same exponent $-\alpha$ independent of $m$ which we also confirmed numerically. We used this scaling for the coefficients $p_r$ of the gap in the antiferromagnetic case.

\section{Wynn extrapolation and scaling analysis}
\label{App:C}
\begin{figure}[!ht]
	\centering
	 \includegraphics[width=.9\columnwidth]{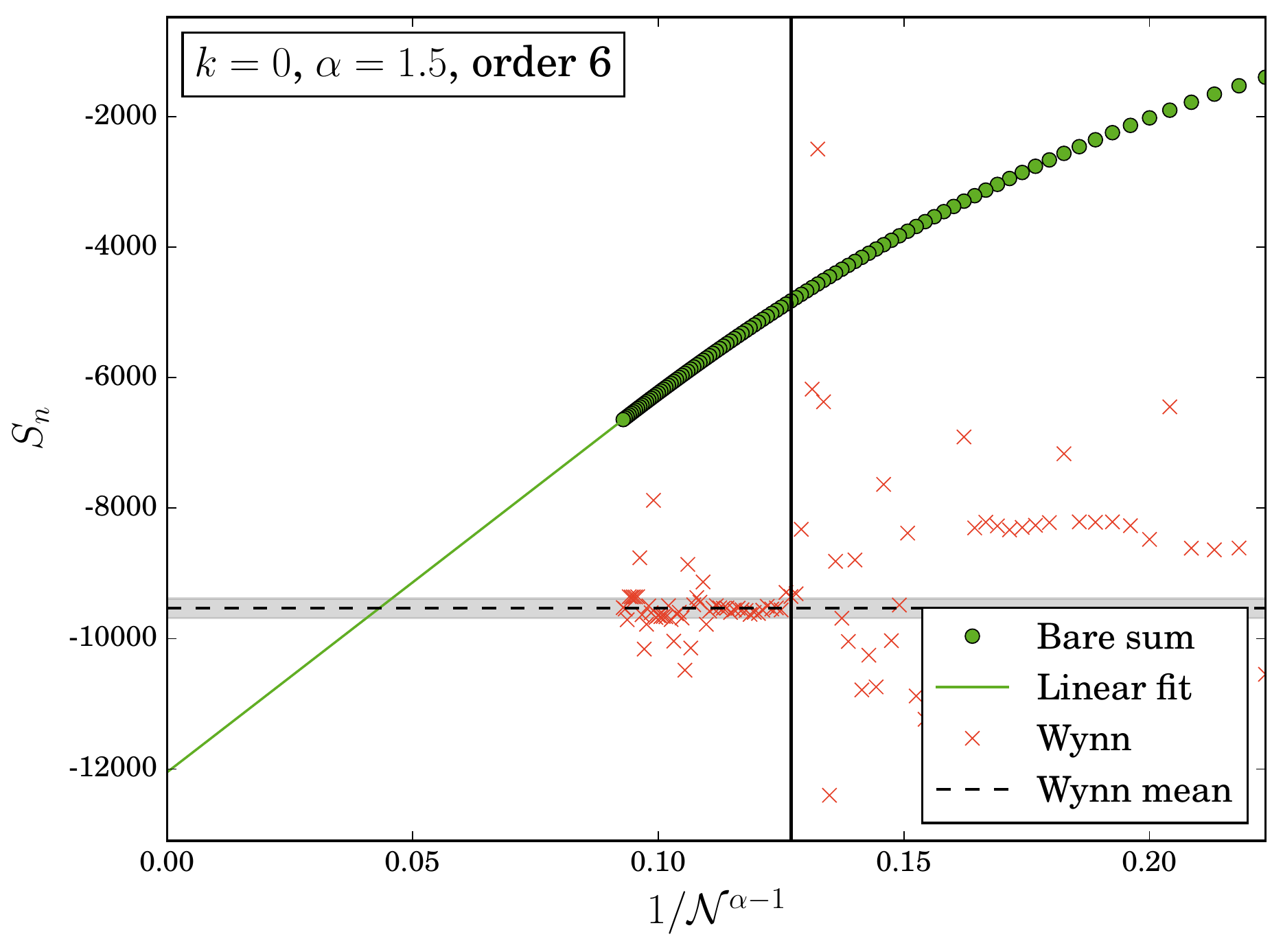}
	 \includegraphics[width=.9\columnwidth]{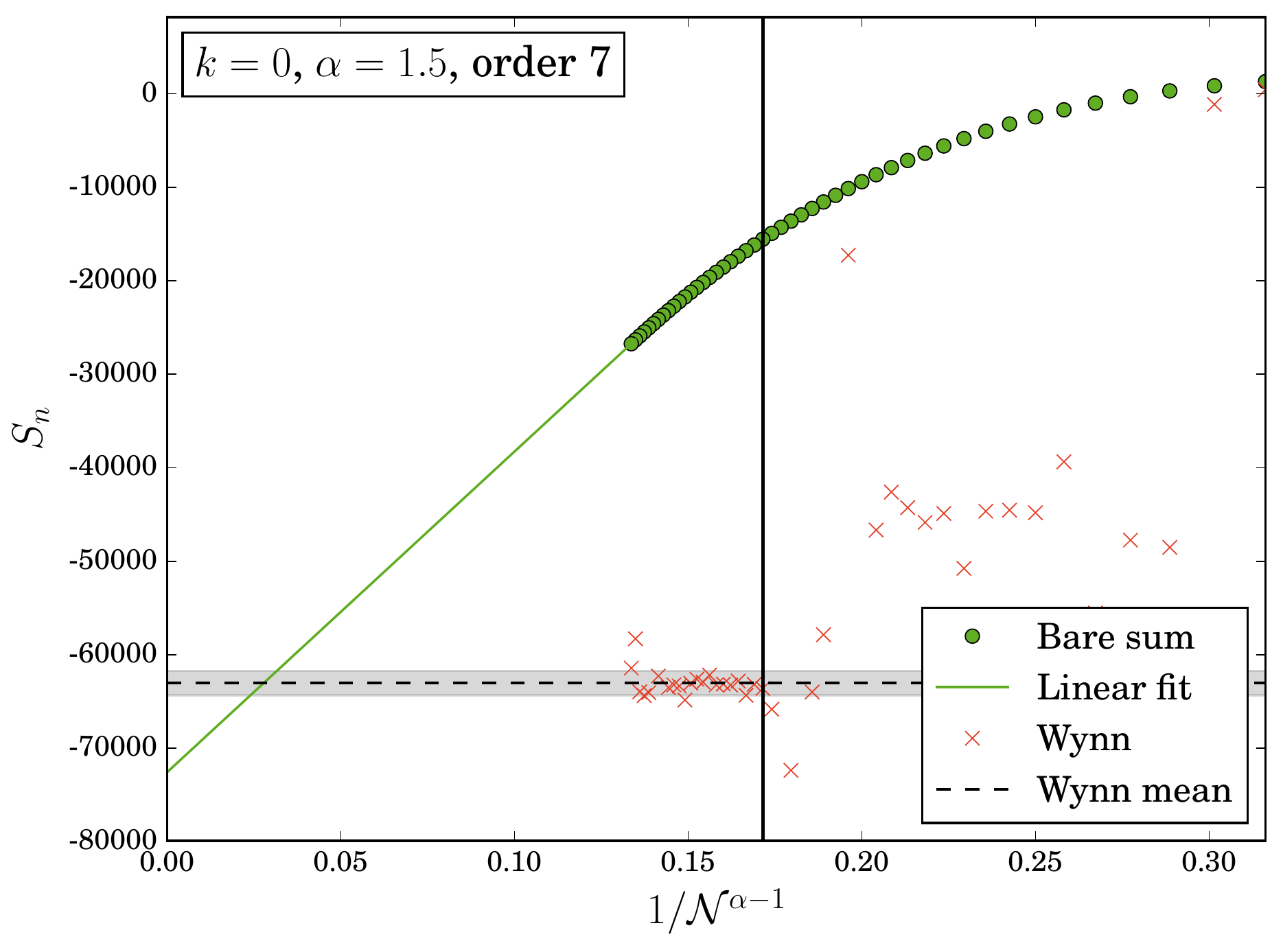}	 
	 \includegraphics[width=.9\columnwidth]{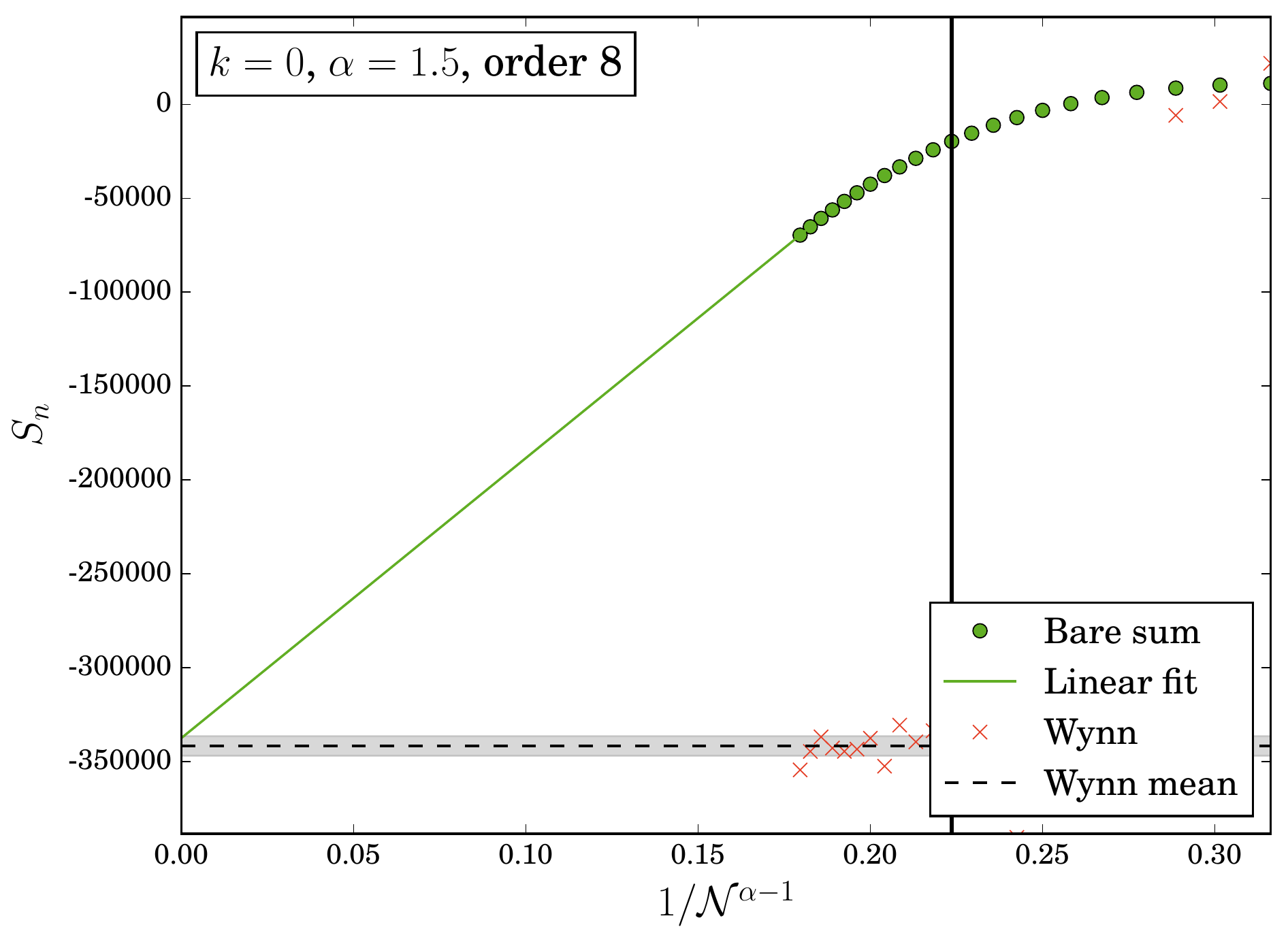}
	 \caption{Wynn extrapolation \& fit for the highest-order prefactors in the ferromagnetic case for $\alpha=1.5$. The black vertical line marks the point after which Wynn extrapolation points are used for calculating the average (dashed black line). The gray area around the mean refers to the standard deviation of those Wynn points.}
	 \label{fig:ext_k0_a1.5}
\end{figure}

\begin{figure}[!ht]
	\centering
	 \includegraphics[width=.9\columnwidth]{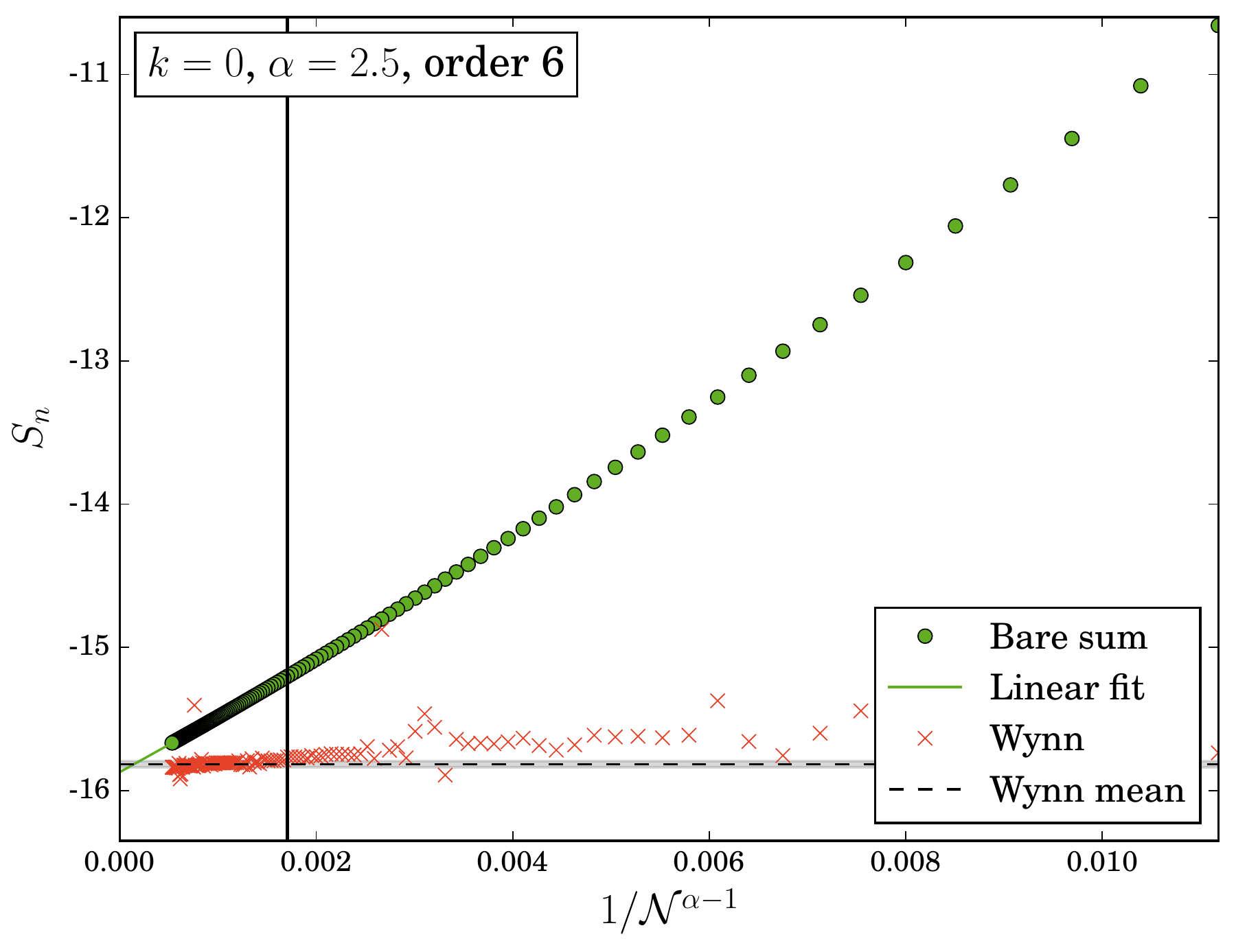}	 
	 \includegraphics[width=.9\columnwidth]{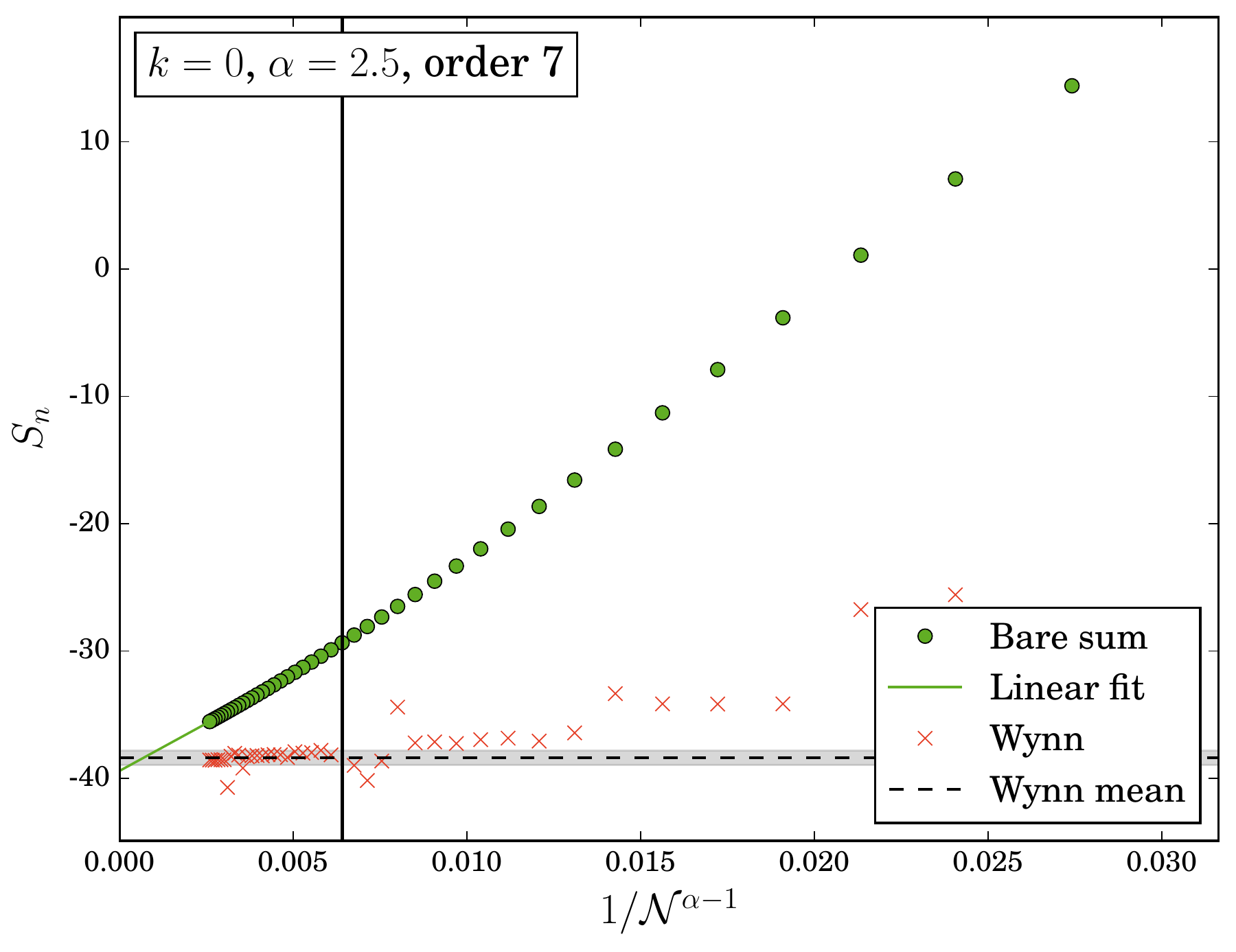}
	 \includegraphics[width=.9\columnwidth]{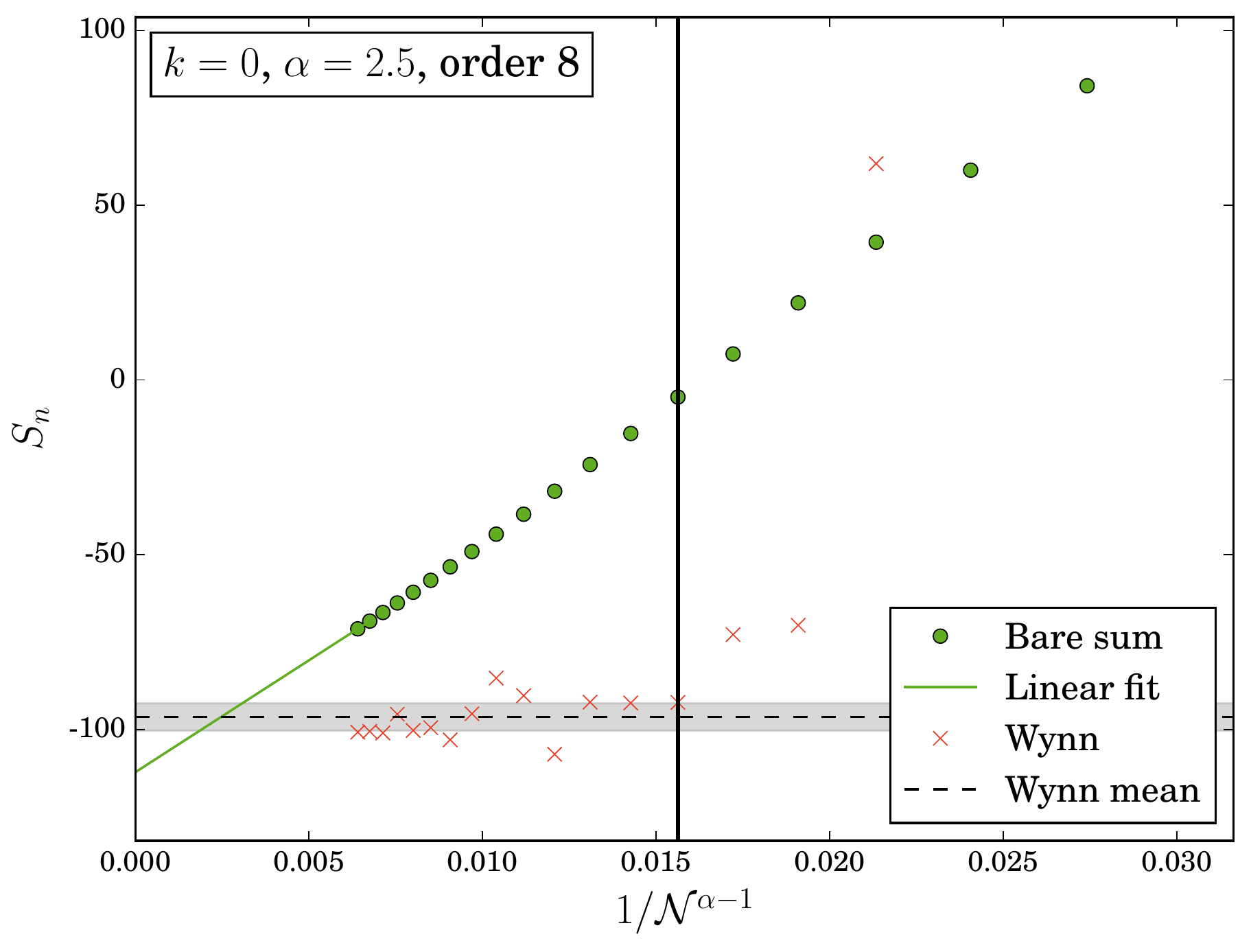}
	 \caption{Wynn extrapolation \& fit for the highest-order prefactors in the ferromagnetic case for $\alpha=2.5$. The black vertical line marks the point after which Wynn extrapolation points are used for calculating the average (dashed black line). The gray area around the mean refers to the standard deviation of those Wynn points.}
	 \label{fig:ext_k0_a2.5}
\end{figure}

\begin{figure}[!ht]
	\centering
	 \includegraphics[width=.9\columnwidth]{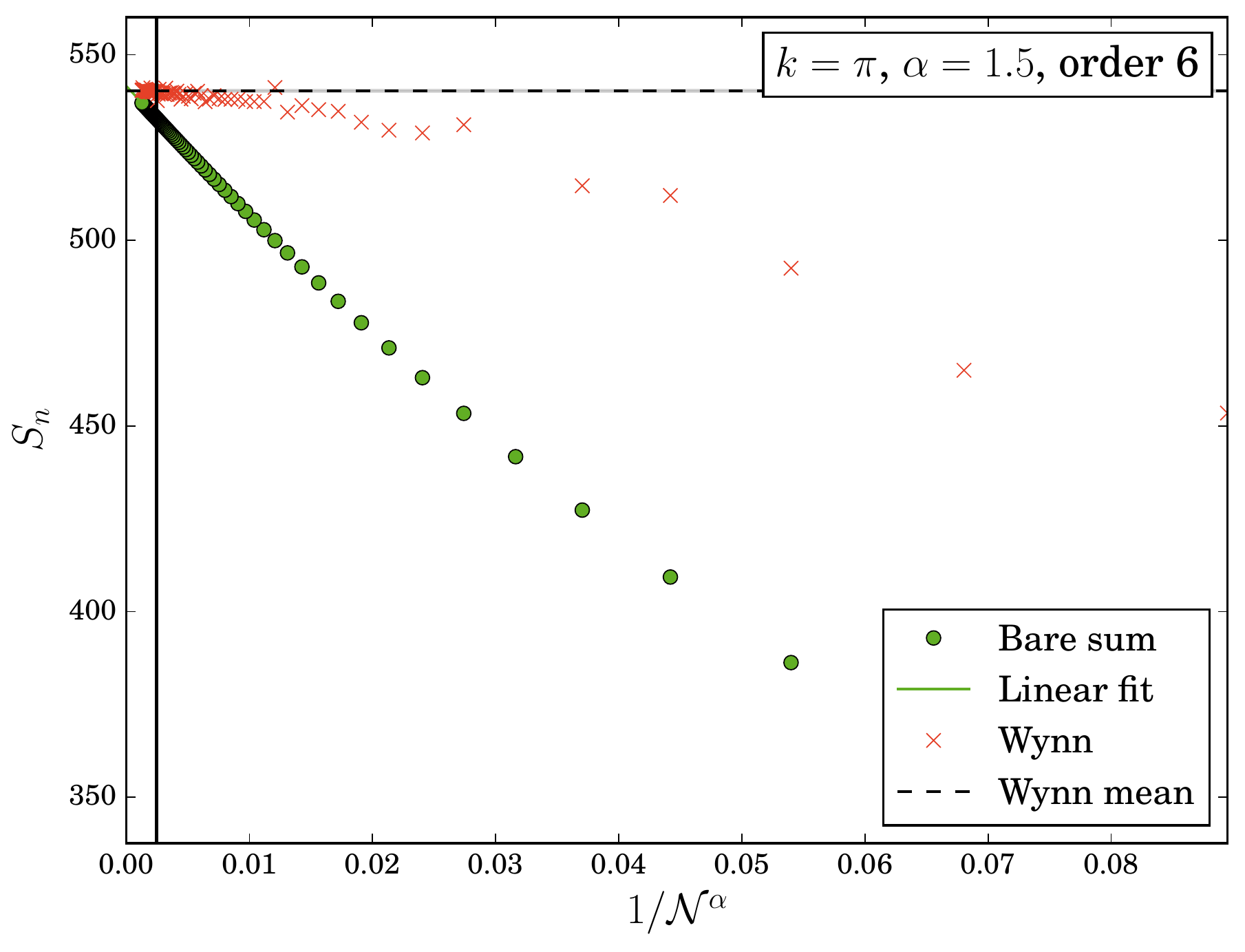}
	 \includegraphics[width=.9\columnwidth]{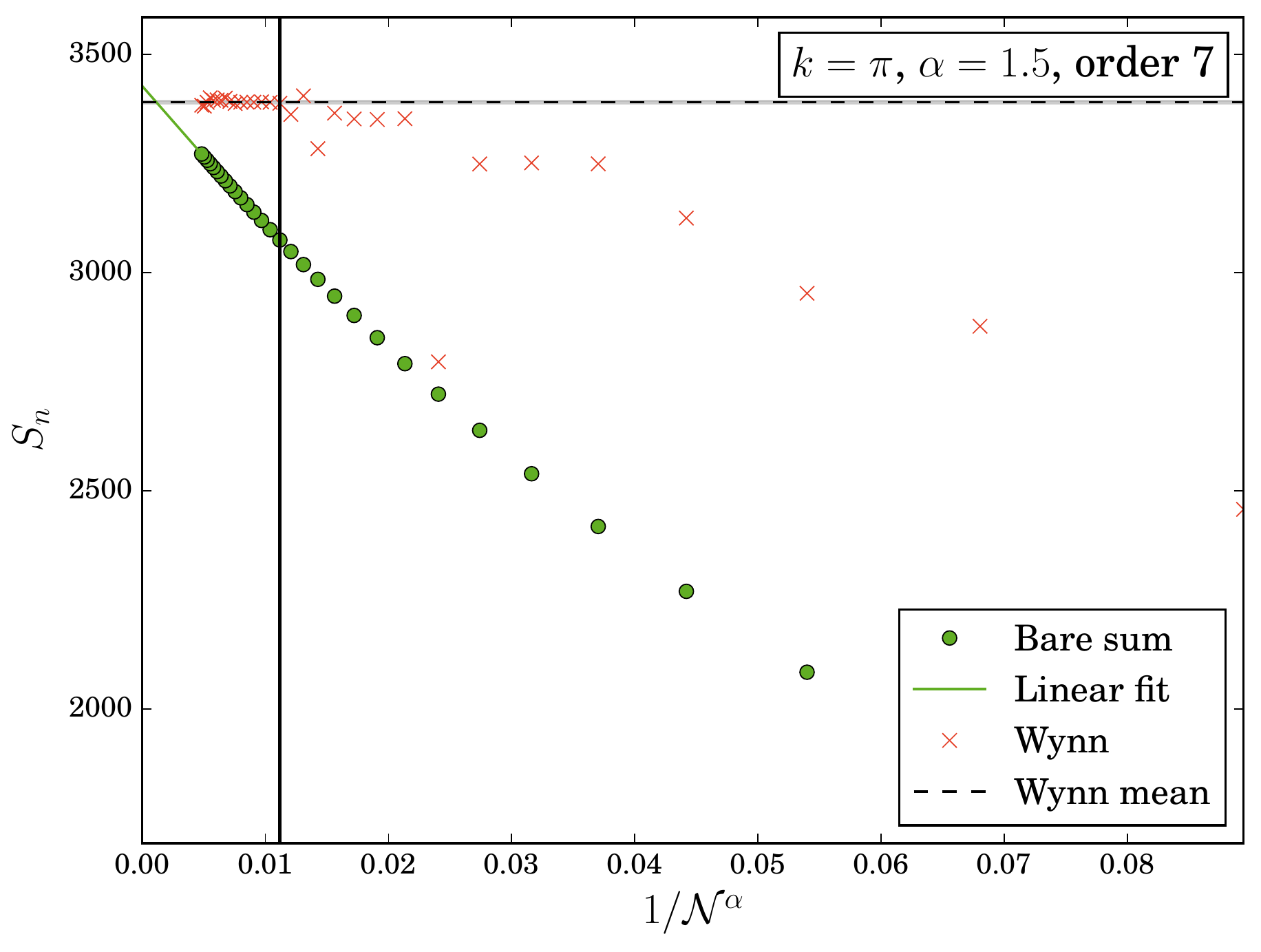}
	 \includegraphics[width=.9\columnwidth]{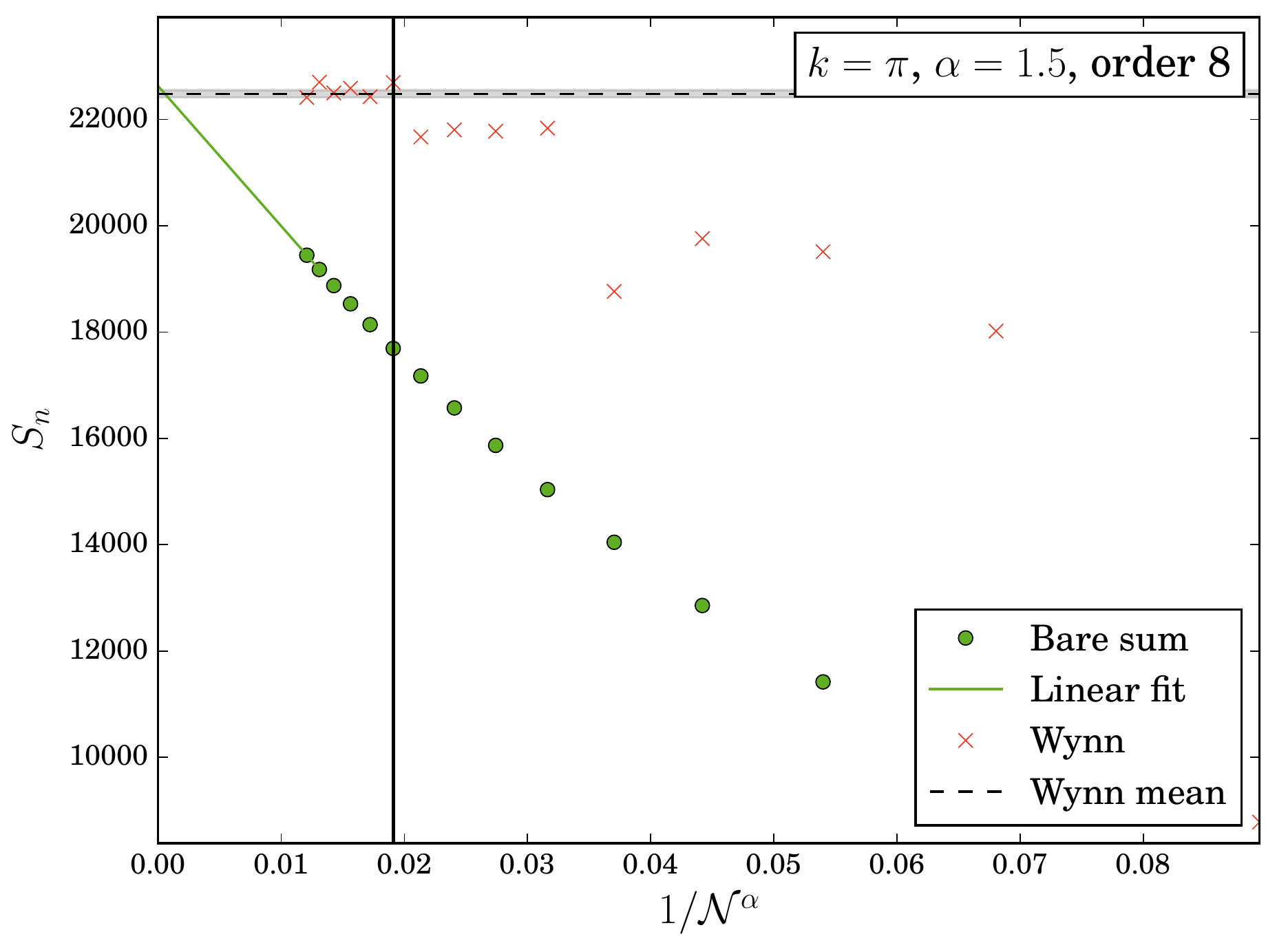}
	 \caption{Wynn extrapolation \& fit for the highest-order prefactors in the antiferromagnetic case for $\alpha=1.5$. The black vertical line marks the point after which Wynn extrapolation points are used for calculating the average (dashed black line). The gray area around the mean refers to the standard deviation of those Wynn points.}
	 \label{fig:ext_kpi_a1.5}
\end{figure}

\begin{figure}[!ht]
	\centering
	 \includegraphics[width=.9\columnwidth]{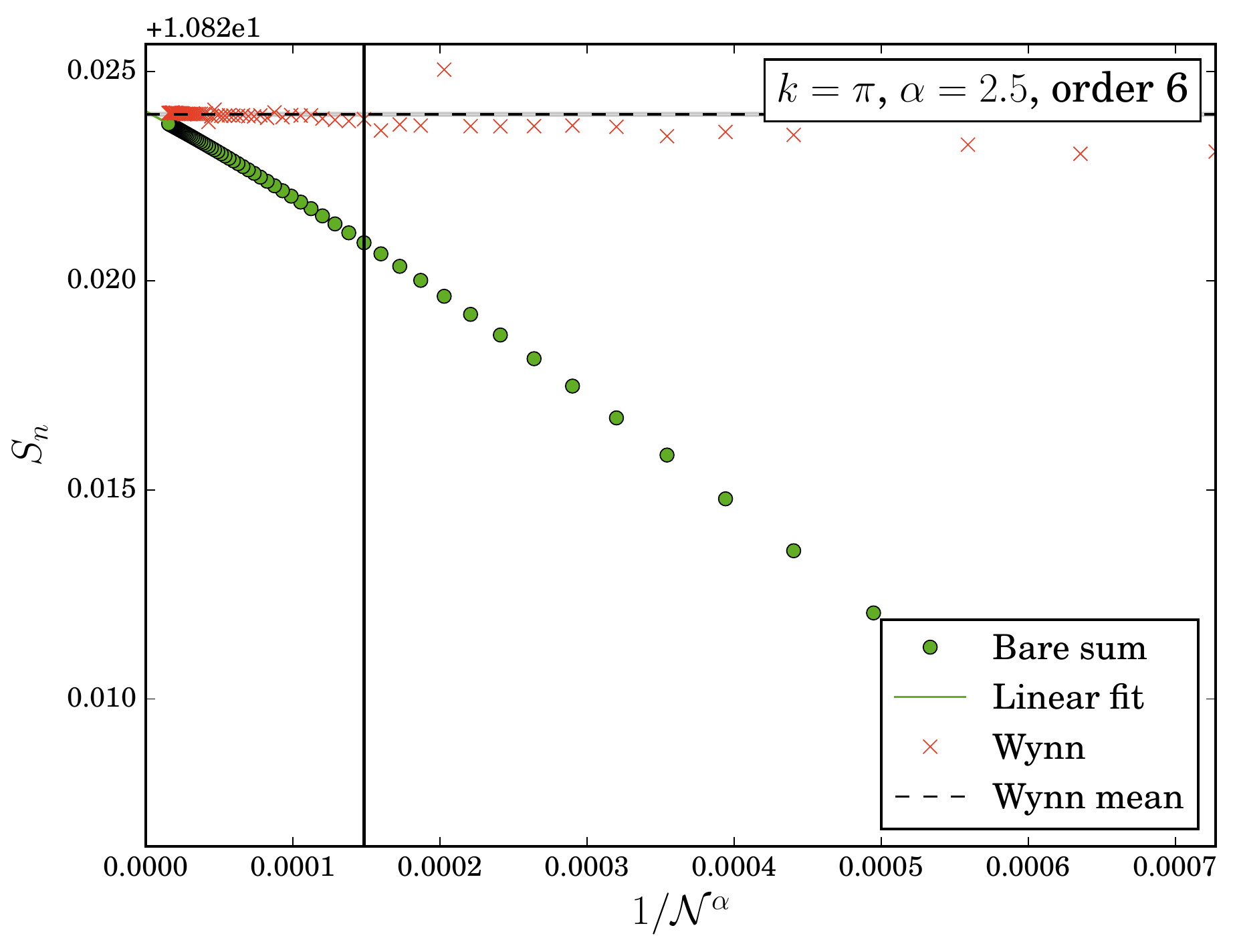}	 
	 \includegraphics[width=.9\columnwidth]{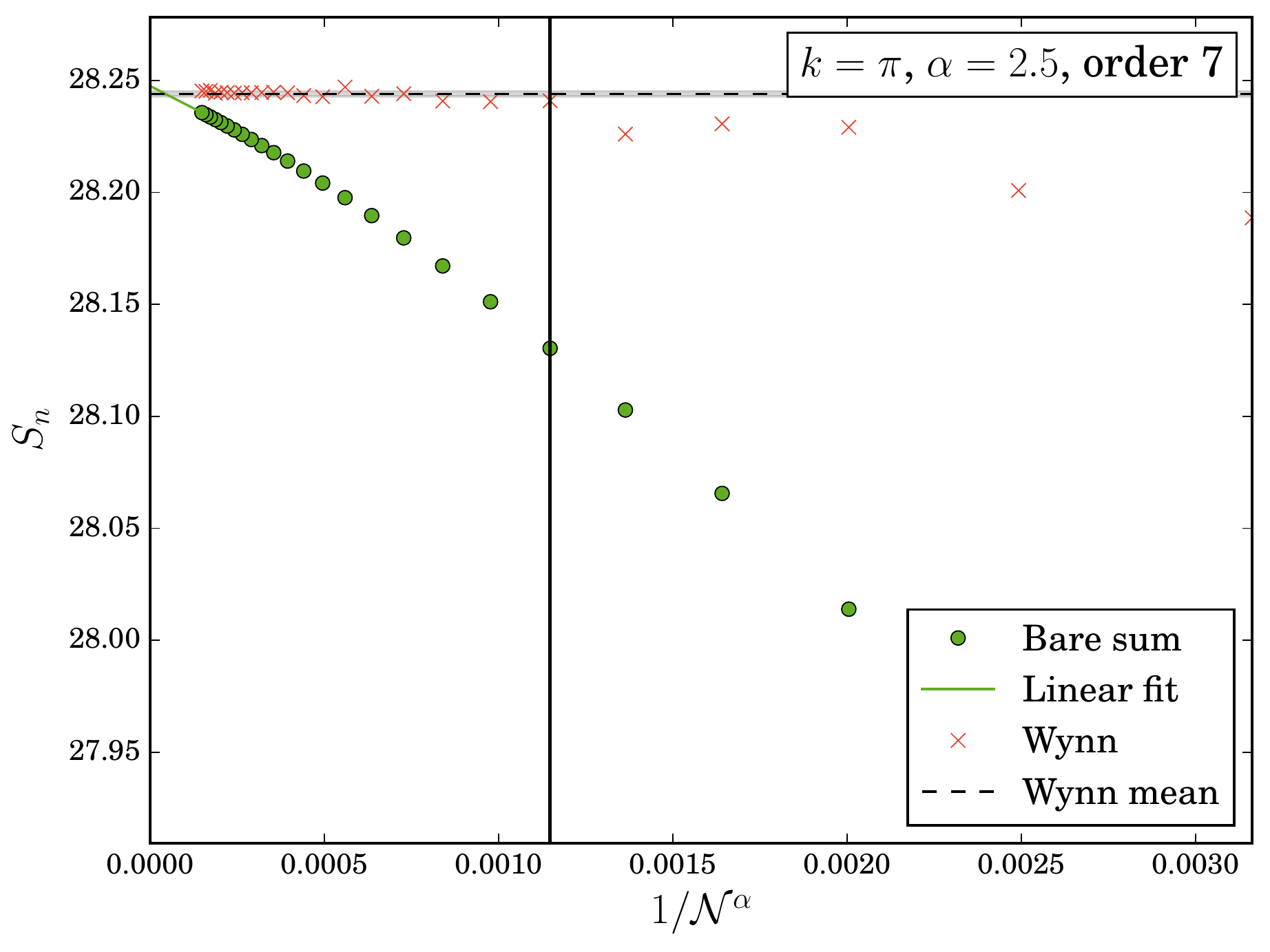}	 
	 \includegraphics[width=.9\columnwidth]{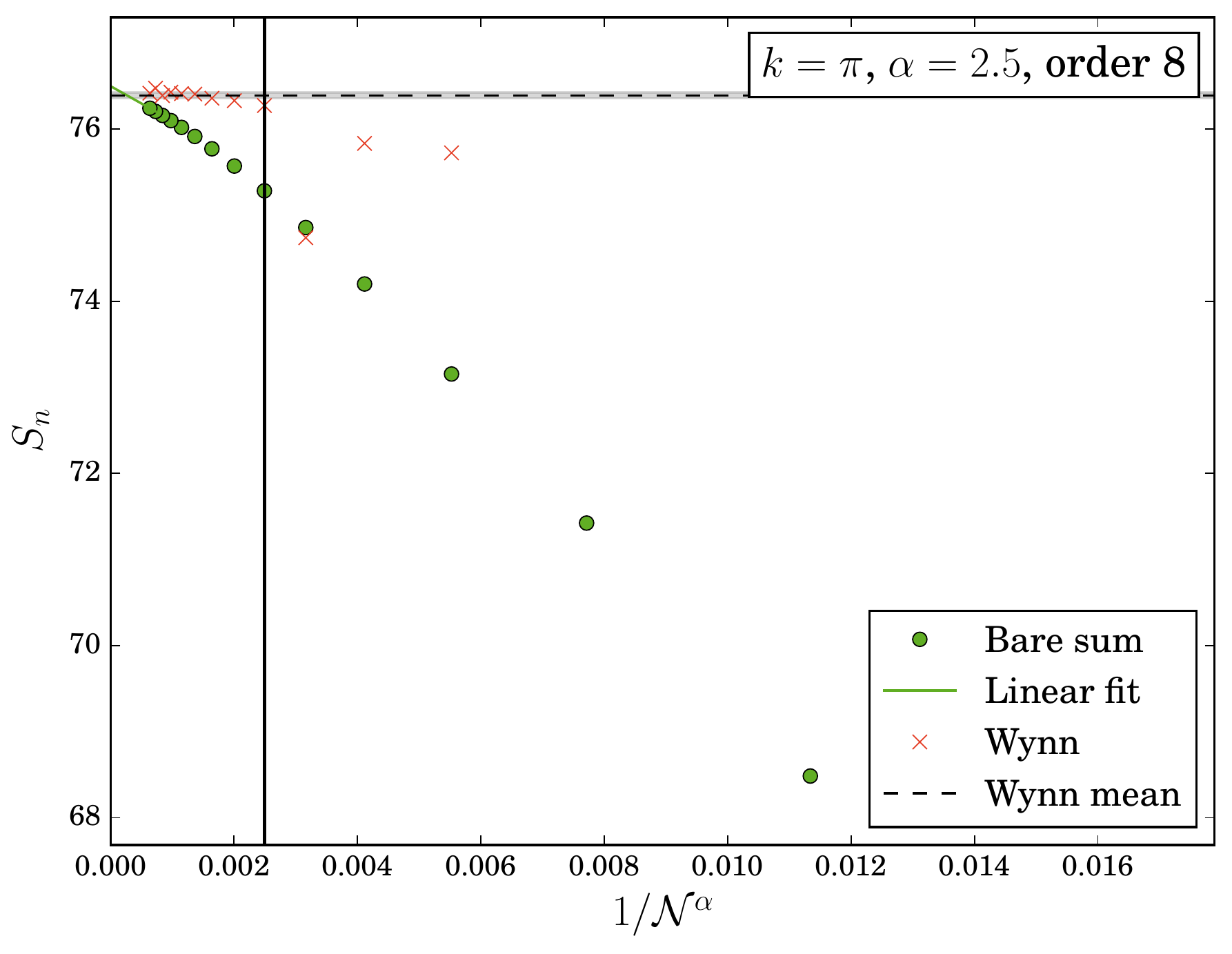}
	 \caption{Wynn extrapolation \& fit for the highest-order prefactors in the antiferromagnetic case for $\alpha=2$. The black vertical line marks the point after which Wynn extrapolation points are used for calculating the average (dashed black line). The gray area around the mean refers to the standard deviation of those Wynn points.}
	 \label{fig:ext_kpi_a2}
\end{figure}

This section contains an exemplary overview of the extrapolations and scalings of the prefactors $p_r$ (c.f. \eqref{eq_supp:gap_series_general}) for both, a ferromagnetic and an antiferromagnetic Ising interaction. Representative data for $\alpha=3/2$ and $\alpha=5/2$ are shown in Figs.~\ref{fig:ext_k0_a1.5} to \ref{fig:ext_kpi_a2}\,\,\,for the highest orders 6, 7, and 8. The contributions from all relevant graphs that are given as nested sums are evaluated up to an upper boundary $\mathcal{N}$ which is only limited by computation time. These partial sums $S_n$ are shown as green circles in the figures.

They are plotted against $n=\frac{1}{\mathcal{N}^{\alpha-1}}$ ($n=\frac{1}{\mathcal{N}^\alpha}$) for a ferromagnetic (antiferromagnetic) Ising interaction. As derived in the previous section the series of points then should display a linear behavior for large $\mathcal{N}$. The last two points (corresponding to the largest $\mathcal{N}$) are used to define a linear curve which gives an estimation for the value of the prefactor for $\mathcal{N}\to\infty$. The curve is shown as a solid green line.

For the calculation of the Wynn extrapolants a subset of partial sums $(S_1,\dots,S_{\mathcal{N}})$ is used and shown as red crosses in the figures. The antiferromagnetic series display an alternating behavior due to the location of the gap at $q=\pi$ (see Eq.~\eqref{eq:embedding}). Only every second value is used to obtain a monotonically converging series. While they give the general tendency, they deviate from the scaled result considerably when looking at small values of $\alpha$ in the ferromagnetic case. However, we found that the differences between the two extrapolation/scaling schemes do influence the final results for the critical values and exponents only marginally.

For a better comparison of Wynn extrapolation and scaling value the Wynn results are averaged from a minimum $\mathcal{N}$ when they seem to have converged. This minimum $\mathcal{N}$ is illustrated by a vertical solid black line in the figures \ref{fig:ext_k0_a1.5} to \ref{fig:ext_kpi_a2}. The standard deviation of these points is illustrated by a gray area.

It can be clearly seen that the prefactors for the antiferromagnetic interaction converge much faster than their ferromagnetic counterpart. Also, as a result, they are in much better agreement with the Wynn extrapolations.


\section{Extrapolation of high-order series}
\label{App:D}
Once the energy gap is given as a power series (c.f. Eq. \eqref{eq_supp:gap_series_general}), we perform standard dLog-Pad\'{e} extrapolations. We refer to the literature for general review of this topic, as for example given in Ref.~\onlinecite{Guttmann1989}. Here we give specific information which is relevant for the particular extrapolation we performed in the main body of the manuscript, which is essentially the information given in Ref.~\onlinecite{Coester2016}. 

Our series are all of the form
\begin{align}
F(\lambda)=\sum_{n\geq 0}^k a_n \lambda^n=a_0+a_1\lambda+a_2\lambda^2+\dots a_k\lambda^k,
\end{align}
with $\lambda\in \mathbb{R}$ and $a_i \in \mathbb{R}$. If one has power-law behavior near a critical value $\lambda_{\rm c}$, the true physical function $\tilde{F}(\lambda)$ close to $\lambda_{\rm c}$ is given by
\begin{align}
\tilde{F}(\lambda)\approx \left(1-\frac{\lambda}{\lambda_{\rm c}}\right)^{-\theta} A(\lambda),
\end{align}
where $\theta$ is the associated critical exponent. If $A(\lambda)$ is analytic at $\lambda=\lambda_{\rm c}$, we can write
\begin{align}
\label{eq:Ftilde}
\tilde{F}(\lambda)\approx \left(1-\frac{\lambda}{\lambda_{\rm c}}\right)^{-\theta}A|_{\lambda=\lambda_{\rm c}}\left(1+\mathcal{O}\left(1-\frac{\lambda}{\lambda_{\rm c}}\right)\right).
\end{align}
Near the critical value $\lambda_{\rm c}$, the logarithmic derivative is then given by
\begin{align}
\tilde{D}(\lambda)&:=\frac{\text{d}}{\text{d}\lambda}\ln{\tilde{F}(\lambda)}\label{dx}\\
&\approx \frac{\theta}{\lambda_{\rm c}-\lambda}\left\{ 1+ \mathcal{O}(\lambda-\lambda_{\rm c})\right\}\nonumber.
\end{align}
In the case of power-law behavior, the logarithmic derivative $\tilde{D}(\lambda)$ is therefore expected to exhibit a single pole at $\lambda\equiv\lambda_{\rm c}$.

The latter is the reason why so-called Dlog-Pad\'{e} extrapolation is often used to extract critical points and critical exponents from high-order series expansions. Dlog-Pad\'e extrapolants of $F(\lambda)$ are defined by
\begin{align}
\label{eq:dlogP1}
dP[L/M]_F(\lambda)=\exp\left(\int_{0}^\lambda P[L/M]_{D}\,\,\text{d}\lambda'\right)
\end{align}
and represent physically grounded extrapolants in the case of a second-order phase transition. Here $P[L/M]_{D}$ denotes a standard Pad\'e extrapolation of the logarithmic derivative
\begin{align}
\label{eq:dlogP2}
P[L/M]_{D}:=\frac{P_L(\lambda)}{Q_M(\lambda)}=\frac{p_0+p_1\lambda+\dots + p_L \lambda^L}{q_0+q_1\lambda+\dots q_M \lambda^M}\quad,
\end{align}
with $p_i\in \mathbb{R}$ and $q_i \in \mathbb{R}$ and $q_0=1$. Additionally, $L$ and $M$ have to be chosen so that $L+M-1\leq k$. Physical poles of $P[L/M]_{D}(\lambda)$ then indicate critical values $\lambda_{\rm c}$ while the corresponding critical exponent of the pole $\lambda_{\rm c}$ can be deduced by
\begin{align}
\theta\equiv\left.\frac{P_L(\lambda)}{\frac{\text{d}}{\text{d}\lambda} Q_M(\lambda)}\right|_{\lambda=\lambda_{\rm c}} \label{extract_exponent}.
\end{align}
If the exact value (or a quantitative estimate from other approaches) of $\lambda_{\rm c}$ is known, one can obtain better estimates of the critical exponent by defining
\begin{align*}
\theta^*(\lambda)&\equiv(\lambda_{\rm c}-\lambda)D(\lambda)\\
&\approx \theta+\mathcal{O}(\lambda-\lambda_{\rm c}),
\end{align*}
where $D(\lambda)$ is given by Eq.~\eqref{dx}. Then
\begin{align}
P[L/M]_{\theta^*}\big|_{\lambda=\lambda_{\rm c}}=\theta \label{biasnue}
\end{align}
yields a (biased) estimate of the critical exponent.

In the ferromagnetic case at the upper critical $\alpha=5/3$, the long-range TFIM displays multiplicative corrections close to the quantum critical point so that one expects the following critical behavior
\begin{align}
\bar{F}(\lambda)\approx \left(1-\frac{\lambda}{\lambda_{\rm c}}\right)^{-\theta} \left(\ln\left( 1-\frac{\lambda}{\lambda_{\rm c}}\right)\right)^{p} \bar{A}(\lambda),
\end{align}
where $\lambda_{\rm c}$ ($\theta$) is the associated critical point (exponent) as before while $p$ yields the exponent of multiplicative logarithmic corrections. Clearly, the extraction of $p$ from a high-order series expansion is very demanding. The only reasonable approach is to bias the extrapolation by fixing $\theta$. In our case the critical exponent $\theta$ is given by the well-known mean-field value $1/2$.

Assuming again that the function $\bar{A}(\lambda)$ is analytic close to $\lambda_{\rm c}$, Eq.~\eqref{eq:Ftilde} transforms into
\begin{eqnarray}
\label{eq:Fbar}
\bar{F}(\lambda)&\approx& \left(1-\frac{\lambda}{\lambda_{\rm c}}\right)^{-\theta} \left(\ln\left( 1-\frac{\lambda}{\lambda_{\rm c}}\right)\right)^{p}\bar{A}|_{\lambda=\lambda_{\rm c}}\nonumber\\
           && \cdot\left(1+\mathcal{O}\left(1-\frac{\lambda}{\lambda_{\rm c}}\right)\right).
\end{eqnarray}
and the logarithmic derivative Eq.~\eqref{dx} becomes
\begin{align}
\bar{D}(\lambda)&\approx \frac{\theta}{\lambda_{\rm c}-\lambda} + \frac{-p}{\ln\left(1-\lambda/\lambda_{\rm c}\right)\left(\lambda_{\rm c}-\lambda\right)} + \mathcal{O}\left(\lambda-\lambda_{\rm c}\right)\nonumber.
\end{align}
One can then estimate the multiplicative logarithmic correction $p$ by defining
\begin{align*}
  p^{*}(\lambda)&\equiv -\ln\left( 1-\lambda/\lambda_{\rm c}\right) \left[  \left( \lambda_{\rm c}-\lambda\right) D(\lambda)-\theta \right]\\
           &\approx p +\mathcal{O}(\lambda-\lambda_{\rm c}),
\end{align*}
and by performing Pad\'{e} extrapolants of this function
\begin{align}
P[L/M]_{p*}\big|_{\lambda=\lambda_{\rm c}}=p \label{biasp}\quad .
\end{align}

\end{document}